\documentclass[]{aa} 

\usepackage{graphicx}
\usepackage{txfonts}
\usepackage{natbib}
\usepackage{url}
\usepackage{epstopdf} 
\usepackage[figuresright]{rotating}
\newcommand{\mygi}{MyGIsFOS}
\newcommand{\Teff}{\ensuremath{T_\mathrm{eff}}}
\newcommand{\g}{\ensuremath{g}}

\newcommand{\glog}{\ensuremath{\log\g}}

\newcommand{\tauross}{\ensuremath{\tau_{\mathrm{ross}}}}

\newcommand{\fei}{\ion{Fe}{i}}
\newcommand{\feii}{\ion{Fe}{ii}}
\newcommand{\alphafe}{[$\alpha$/Fe]}
\newcommand{\Vturb}{V$_\mathrm{turb}$}
\newcommand{\feh}{[Fe/H]}

\begin{document}

   \title{MyGIsFOS: an automated code for parameter determination and detailed abundance analysis in cool stars}

   \author{L. Sbordone\inst{1,2}
          \and
          E. Caffau\thanks{Gliese Fellow}\inst{1,2}
          \and
          P. Bonifacio\inst{2}
          \and
          S. Duffau\inst{1}
          }

   \institute{	
	{Zentrum f\"ur Astronomie der Universit\"at Heidelberg, Landessternwarte, K\"onigstuhl 12, 69117 Heidelberg, Germany}
	\\
	\email{lsbordon@lsw.uni-heidelberg.de}
	\and
	{GEPI, Observatoire de Paris, CNRS, Universit\'e Paris Diderot ; Place Jules Janssen, 92190 Meudon, France}
             }

\authorrunning{Sbordone et al.}
\titlerunning{A code for automatic abundance analysis}
\date{Received ...; accepted ... }

  \abstract
   {The current and planned high-resolution, high-multiplexity stellar spectroscopic surveys, as well as the swelling amount of under-utilized data present in  public archives have led to an increasing number of efforts to automate the crucial but slow process to retrieve stellar parameters and chemical abundances from spectra.}
   {We present \mygi, a code designed to derive atmospheric parameters and detailed stellar abundances from medium - high resolution spectra of cool (FGK) stars. We describe the general structure and workings of the code, present analyses of a number of well studied stars representative of the parameter space \mygi\ is designed to cover, and examples of the exploitation of \mygi\ very fast analysis to assess uncertainties through Montecarlo tests.}
   {\mygi\ aims to reproduce a ``traditional'' manual analysis by fitting spectral features for different elements against a precomputed grid of synthetic spectra. \fei\  and \feii\ lines 
can be employed to determine temperature, gravity, microturbulence, and metallicity by iteratively minimizing the dependence of \fei\ abundance from line lower energy and equivalent width, and imposing \fei-\feii\ ionization equilibrium. Once parameters are retrieved, detailed chemical abundances are measured from lines of other elements.}
   {\mygi\ replicates closely the results obtained in similar analyses on a set of well known stars. It is also quite fast, performing a full parameter determination and detailed abundance analysis in about two minutes per star on a mainstream desktop computer. {Currently, its preferred field of application are high-resolution and/or large spectral coverage data (e.g UVES, X-Shooter, HARPS, Sophie).}}
   {}


   \maketitle
%

\section{Introduction}

The availability of several high-efficiency, high-multiplexity spectrographs has 
brought about the need to perform accurate abundance analysis on 
large sets of stellar spectra of low to high resolution.
The problem has been tackled in many different ways, 
one may roughly divide the methods into ``global'', that make use
of the whole spectrum \citep[e.g.][]{katz98,A00,BJ2000,recio06} and 
``local'', that make use only of selected sections
of the spectrum \citep[e.g.][]{erspnorth,bonifacio03,Bark05,boeche,spades,gala,fama}.
{A few ``intermediate'' cases exist, most notably SME \citep{valenti96,Bark05}, underlining 
perhaps the difficulty to come up with a clear-cut classification scheme.}
Complex pipelines, like that of the Sloan Digital Sky Survey \citep{A08}
use multiple methods whose results are then suitably combined.
Among the ``local'' codes one may distinguish between those that rely
on equivalent widths (EWs) measured by automatic codes such as
{\tt fitline} \citep{francois},
DAOSPEC \citep{stetson08} or ARES \citep{ares} 
{to} determine stellar parameters and abundances \citep{gala,fama}, and those that 
rely on line-fitting \citep{erspnorth,bonifacio03,Bark05,boeche,spades,vds}.
\citet{A04} argued that EW based analysis should be abandoned, 
see also \citet{bonifacio05} on EWs and line-fitting.

We present in this paper the code \mygi\ that uses a ``local'' approach
to treat large numbers of medium to high-resolution spectra. 
{Among the ``local'' codes \mygi\,  Abbo \citep{bonifacio03}, and the RAVE pipeline of \citet{boeche}
are the only ones, to our knowledge, that do not perform ``on-the-fly'' line transfer computations, 
but rely {\em only} on a pre-computed grid. In the \citet{boeche} pipeline, a library of synthetic curves of
growth are compared to the measured EW of the chosen observed lines. In this sense, this code is closer to
EW-based ones as far as line selection criteria, advantages, and limitations are concerned, but faster than most EW-based codes
since no on-the-fly line transfer is performed. 

\mygi\ and Abbo, on the other hand, directly fit the {\em synthetic profile} of each chosen feature against the 
observed one. As such, they allow to circumvent some of the limitations of EW-based codes (see, e.g., Sect. \ref{featvslines}), while, at the same time, 
maintaining the speed advantage of codes not performing on-the-fly calculations.

At the same time, the RAVE pipeline, \mygi\ and Abbo suffer of limitations inherent to the use of pre-computed grids. Namely, these grid can become 
exceedingly large, or must be limited in parameter space range. Moreover, precomputed grids are by their nature ``rigid'': their computation
is resource intensive, and recomputation for the purpose of changing, for instance, one of two line oscillator strength might not be 
a desirable option.}

{For all these reasons, such codes} are optimized to analyze a large number of stars that span
a limited range in metallicity, effective temperature and surface gravity.

{\mygi\ specifically targets the treatment of large amounts of data, through a local approach,
based on spectrum synthesis,  line profile fitting and a pre-computed synthetic spectrum grid.
Although other procedures exist that incorpoarate these feature none exists that uses at the
same time all of these features. In this case we believe \mygi\ is unique and innovative.}


\section{The purpose of \mygi}
\mygi~ is  built on the foundation of our previous automatic abundance analysis code Abbo \citep[][]{bonifacio03}. Although it has been completely rewritten, has a different input-output system and it is considerably more powerful and faster, the scope of the code is unchanged with respect to Abbo. Broadly speaking, the observed spectrum is compared against a suitable grid of synthetic stellar spectra which have been computed at varying \Teff, \glog, \Vturb, [Fe/H], \alphafe. Selected \fei, \feii, and $\alpha$-elements features are used to iteratively estimate the best values for each parameter, after which features for other elements are fitted to derive the corresponding abundances.

\begin{table*}
\caption{The \mygi\ grids {computed with SYNTHE and ATLAS 12 models} employed for the present paper.}             
\label{gridlist}      
\centering                  
{
\begin{tabular}{l l l l l l l}        
\hline\hline                 
Grid  & \Teff\  & \glog\ & \Vturb\ & \feh\ & \alphafe & Number of \\
name  & range   & range  & range   & range & range    & models    \\
      & [K]     & [c.g.s.] & [ km s$^{-1}$] \\
\hline
\\
Metal Poor Cool Dwarfs & 5000 to 6000 & 3 to 5   & 0 to 3             & -4 to -0.5 & -0.4 to 0.8 & 3840 \\
(MPCD)                 & step 200 K   & step 0.5 & step 1 & step 0.5   & step 0.4    &      \\
\\
Metal Rich Cool Dwarfs & 5000 to 6000 & 3 to 5   & 0 to 3             & -1 to 0.75 & -0.4 to 0.8 & 3840 \\
(MRCD)                 & step 200 K   & step 0.5 & step 1 & step 0.25  & step 0.4    &      \\
\\
Metal Poor Giants      & 4200 to 5600 & 1 to 3   & 1 to 3             & -4 to -0.5 & -0.4 to 0.4 & 2880 \\
(MPG)                  & step 200 K   & step 0.5 & step 1  & step 0.5   & step 0.4    &      \\
\\
Metal Rich Giants      & 4200 to 5600 & 1 to 3   & 1 to 3             & -1 to 0.5 & -0.4 to 0.4 & 2520  \\
(MRG)                  & step 200 K   & step 0.5 & step 1  & step 0.25   & step 0.4  &       \\
\hline                      
\end{tabular}
}
\end{table*}

\mygi~ is conceived to strictly replicate a ``classical'' or ``manual'' procedure to derive stellar atmospheric parameters and detailed chemical abundances from high resolution stellar spectra of cool stars. As such, its most typical usage case can be resumed as follows:

\begin{itemize}
\item {For each star to be analyzed, the user provides input observed spectrum and a set of first guess parameters.}
\item The user provides the code a feature list, i.e. a list of spectral intervals for the grid to be fitted against the observed spectrum. In the most general case the feature list will include continuum intervals for pseudo-normalization and signal-to-noise ratio (S/N) estimation, intervals corresponding to \fei\, \feii\, and $\alpha$-element lines for atmospheric parameter and metallicity ([Fe/H]) determination, and intervals corresponding to lines of other elements for the determination of detailed abundances.
\item \Teff~ is determined from a set of isolated  \fei\ lines imposing the linear fit of the transition lower energy vs. line abundance to have null slope (for brevity, we will henceforth refer to this quantity as Lower Energy Abundance Slope, or {LEAS}).
\item \Vturb\ is determined by imposing null slope for the Equivalent Width (EW) vs. abundance relation of isolated \fei\ lines.
\item \glog\ is determined from \fei - \feii\ ionization equilibrium.
\item Fe abundance is determined from \fei\ lines only.
\item \alphafe\ is determined by measuring lines of various $\alpha$ elements, and using their average [X/Fe] as an estimate of \alphafe.
\item \Vturb, \glog, and \alphafe\ are estimated iteratively, in a nested fashion, \alphafe\ being the outermost ``shell'', \Vturb\ the innermost. For a given set of current guesses of \Teff, \glog, and $\alpha$ enhancement, \Vturb\ is determined, then the code passes to update the gravity estimate: if the current one is not appropriate, a new one is guessed, \Vturb\ is redetermined by assuming the new gravity and the existing estimates for the other parameters, then gravity is tested again. When a satisfactory estimate is reached for both \Vturb\ and \glog, \alphafe\ is tested, and if changed, a recalculation of \Vturb\ and \glog\ is triggered. 
\item \Teff\ is evaluated in a slightly different fashion: initially, the aforementioned analysis is repeated to full convergence assuming a set of different temperatures, and in each case, the LEAS is derived. This intial ``mapping'' is used to fit the \Teff-LEAS relationship and look for a zero-slope temperature, which is then used to repeat the \Vturb, \glog, \alphafe\ determination. The LEAS is evaluated again and the estimated slope added to the \Teff-LEAS relationship fitting sample. A zero is searched again and so on, until convergence is reached. 
\item Any of the aforementioned parameters can be either derived as described from the spectrum, or kept fixed at a user-defined value. \mygi\ will of course refrain to alter parameters for which the corresponding estimator is absent, e.g. gravity will not be estimated if no \feii\ features are successfully measured. Thus, the user does not need to provide features for any parameter he/she is not planning to alter, the only exception being \fei\ features, which should always be present (\feh\ cannot be kept fixed).
\item The whole analysis is executed in a fully automated way for all the stars in the input list 
and the output is stored in a separate directory for each star.
The input star list should contain data of similar quality and spectral range \citep[e.g. UVES red-arm spectra with S/N$\sim$ 50-100, for a description of UVES see][]{dekker00}, and of objects of comparable characteristics (say, metal-rich FGK dwarfs), because feature list, general running parameters (e.g. fit rejection tolerances) and synthetic spectra grid are common to the input star list and should thus be appropriate for all the objects.
{The constraints on S/N ratio, spectral type and metallicity,
are not very  tight. Experiments with real UVES spectra
have shown that if the feature list has been optimized for S/N in the range 50-100 it can works
well for S/N in the ratio 15-300. At the low S/N ratio most of the selected features 
are not detected because too weak and one has to switch to a selection of strong, saturated lines.
At the high S/N ratio end it is useful to add many weak lines
that are not detectable at lower S/N.
For the atmospheric parameters the  main constraint is metallicity, since
features that are heavily blended become much cleaner and are usable at lower
metallicity, at the same time features that are saturated
in the high metallicity regime reach the linear part of the
curve of growth at lower metallicity. 
The switch-over between metal-rich and meal-weak regime happens
somewhere between [M/H] --0.5 and --1.0. \mygi\ is designed to analyse data
sets for which we have some previous knowledge (colours, metallicity estimates,
membership to a cluster or dwarf galaxy) our experience is that 
in these cases the stars that need to be re-run  a second time
because initially misclassified is less than 5\%.
} 
\end{itemize}

It is then clear that the ideal application of \mygi\ is the determination of detailed chemical abundances from high to intermediate
resolution spectra of cool stars, with basically the same limitations and strengths as a traditional ``manual'' analysis: the results will be of higher quality if the spectral coverage is large, if S/N ratio is good, if clean, unblended features are chosen. Few, noisy or anyway unreliable \feii\ lines will make gravity estimation difficult, the quality of the adopted atomic data will impact the precision of any abundance derived, and so on. 
On the other hand, \mygi\ results are immediately comparable with the ones derived from a traditional abundance analysis. Uncertainties, limitations, and dependence from the  assumption made in atmosphere modeling and spectrosynthesis are well known, and researchers in the field are since long used to deal with them. This makes \mygi\ output easy to test, interpret, and compare with the results from previous works, {advantages \mygi\  shares with EW-based automation schemes such as FAMA or GALA}.

\mygi\ produces an extensive output (in ASCII format) for each star to allow for critical examination of the analysis outcome. Included are detailed information on each feature fitted (best fitting line profile, abundance, EW, rejection flags and quality-of-fit estimators), pseudo-normalized input spectra and best-fitting synthetic spectra, as well as averaged abundances and parameters in tabular form, and a full listing of all the input parameters, employed grid characteristics, and code version. Also, all output files from the same run share a timestamp to ease tracking.

   \begin{figure*}
   \centering
            \includegraphics[width=15.5cm]{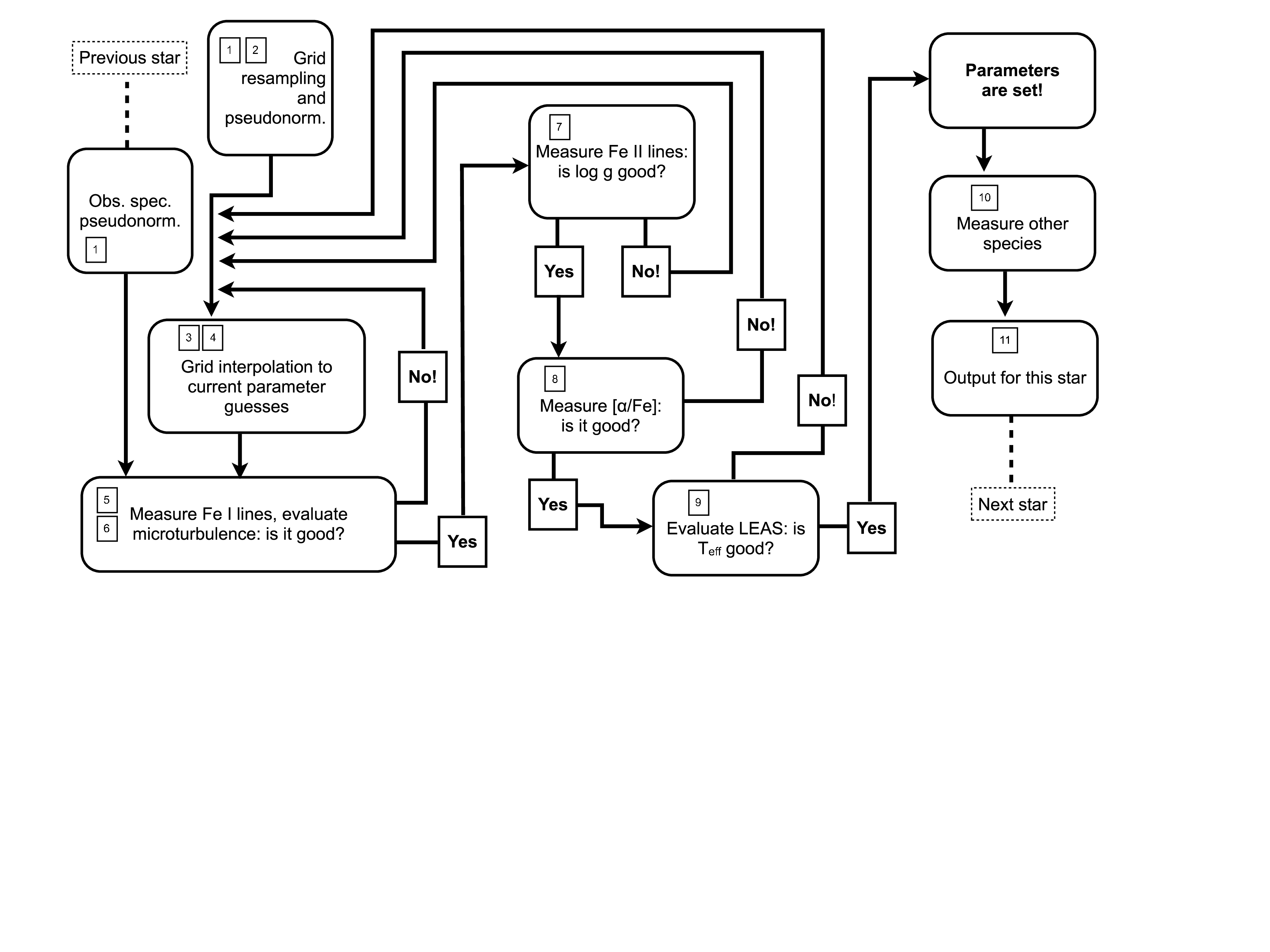}
      \caption{A schematic flowchart of \mygi. {Numbers in rectangles refer to the \mygi\ phases enumeration given in Sect. \ref{workflow}.}}
              
         \label{mygiflowchart}
   \end{figure*}

Thanks to the use of a fully precomputed synthetic grid, \mygi\ is remarkably fast: a typical run on a {standard} desktop computer, with full parameter determination and $\sim$20 element abundances, for 200nm high-resolution optical spectra takes about 120s per star (see Sect. \ref{performa} for more details).


\section{The synthetic spectra grid}
\label{grid}

\mygi~ works by comparing selected spectral features with a grid of synthetic stellar spectra with varying \Teff, \glog, \Vturb, \feh, and \alphafe. The grid is provided as an unformatted binary file that reduces physical size and read-in time.  The grid contains a header containing a comment, the grid ``metrics'' (starting point, step and number of steps in each parameter), the assumed solar abundances, the elements affected by $\alpha$ enhancement and the assumed grid instrumental broadening. The grid should be passed as already broadened as needed by the spectrograph resolution. The grid is in general divided into several spectral ranges or ``frames''. This is foreseen to handle situations in which only limited, non contiguous spectral ranges are needed, such as when treating data from different settings of single-order, high multiplexity spectrographs such as FLAMES@VLT\footnote{Apart from the savings in file size, this allows to apply different instrumental broadening values to each frame as it is needed e.g. for different FLAMES@VLT settings or XSHOOTER@VLT spectral arms.} \citep{pasquini02}. If this is not needed, the grid might contain a single frame covering the needed spectral range. The frame subdivision is also indicated in the grid header, that then contains all the information needed by \mygi~ to read in the grid data without user intervention. 

The grid should be passed to \mygi~ at high sampling ($\Delta \lambda / \lambda >$100000) to prevent the arising of artifacts when it is resampled over the actual observed data points.

   \begin{figure*}
   \centering
\includegraphics[width=8.5cm,angle=90]{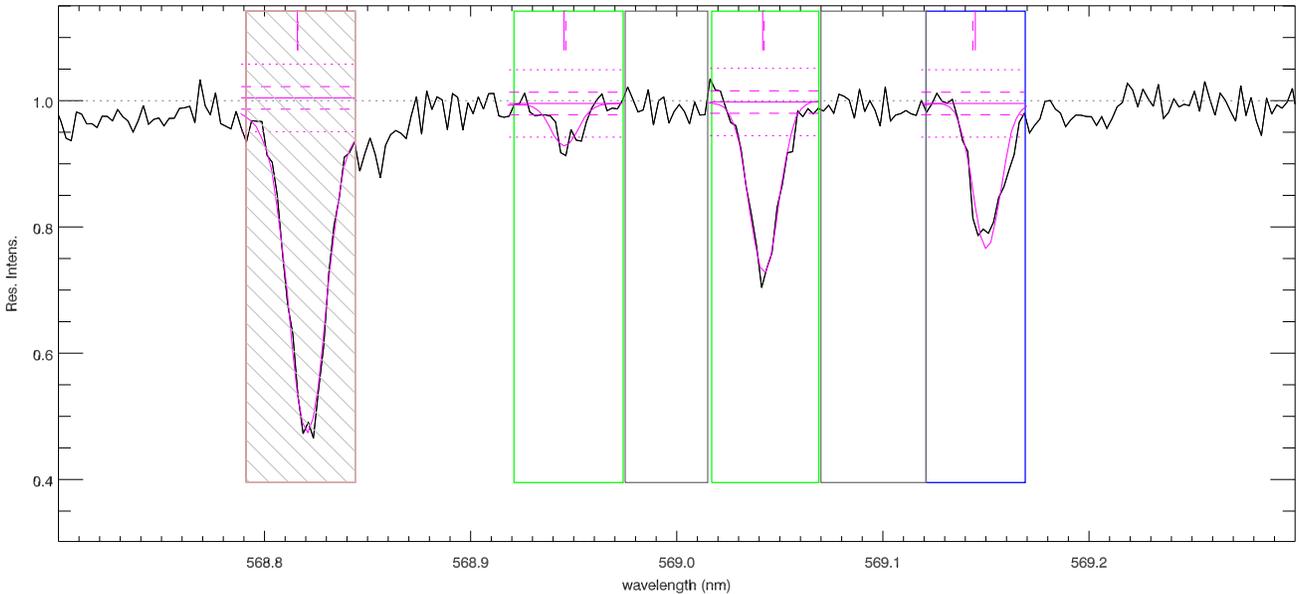}
      \caption{A small section of the S/N=50 solar spectrum presented in \ref{stars:sun} with overplotted \mygi\ features and fitting result, as produced by the ancillary plotting package SQUID. From left to right: a rejected \ion{Na}{i} line (rejected due to EW exceeding the applied maximum EW constraint, gray shaded box); a \ion{Ti}{i} feature (green box); a continuum interval (gray box); a \ion{Si}{i} feature (green box again); another continuum (gray); a \fei\ features (blue box). The observed spectrum (black line) appears as renormalized by \mygi. The gray dashed horizontal line represents the continuum level, each feature is superimposed with the best fit synthetic (magenta), the best fit continuum (magenta continuous horizontal line), 1-$\sigma$ and 3-$\sigma$ values of the noise (horizontal dashed and dotted magenta lines), and markers of best fit doppler shift for the feature (magenta vertical continuous and dashed lines).}\label{squidexamplesun}
   \end{figure*}
   
For all the tests presented in this paper we computed a set of model atmospheres and synthetic spectra appropriate for the analysis of FGK stars, both dwarfs and giants, over a wide range of metallicities. The atmospheric models have been computed with the Linux version of ATLAS 12, while synthetic spectra covering the UVES RED 580 setting (approx 480 to 680 nm) have been computed by means of SYNTHE \citep[][]{kurucz05, castelli05, sbordone04, sbordone05}. 
Parameter space coverage of the used grids to date are listed in Table \ref{gridlist}. Models and syntheses have been computed in a self-consistent way, i. e. changes in \Vturb\ and \alphafe\ have been included already in the opacity computation during model calculation. Assumed solar abundances are derived from \citet{caffau11b} for the elements there analyzed, and from \citet{lodders09} for the remaining species. The used grids are part of a larger set being currently 
prepared for publication \citep{sbordone13}. Synthetic grids are computed with sampling $\Delta \lambda / \lambda$=300000 in the 460nm - 690nm range, and take in their binary packaged version between 1.5 and 2.2 GB of space.


\section{The MyGIsFOS workflow in detail}
\label{workflow}

As above stated, \mygi\ replicates classical ``manual'' abundance analysis workflow. User defined parameters are read from an ASCII parameter file. A flowchart of the process for a single star is shown in Fig.\ref{mygiflowchart},  while an example of the result is given in Fig. \ref{squidexamplesun}. The workflow can be summarized as follows:

\begin{enumerate}
\item The observed spectrum (or spectra) is pseudonormalized by evaluating the pseudo-median of each continuum interval\footnote{Given a vector of $N$ values, the median is defined as the value at the $N/2$-th element of the sorted vector. The pesudo-median employed by \mygi\ is instead the value at the $N/a$-th element of the sorted vector. It is thus equivalent to the true median for $a=2$, but the most typical values employed in\mygi\ are in the range $a=1.25-1.66$: this is done to account for the fact that in most cases the chosen continuum intervals correspond more exactly to pseudo-continua, where  weak lines are buried into the noise so that the use of a straight average or median would underestimate the continuum value.
It is left to the user to fix the proper pseudo-median factor $a$ by visually inspecting the spectra.}. A continuum value is then calculated for each observed pixel by computing a spline through all the continuum intervals. Also, from every continuum interval the local S/N ratio is estimated.
The {whole} synthetic grid is pseudo-normalized 
in the same way. {The pseudonormalization is kept fixed throughout the analysis of the star.}
\item The synthetic grid is then resampled at the wavelengths of each observed pixel.  
\item \label{tcontrol} The first value for effective temperature estimation is chosen. If \Teff\ is kept fixed for the star, said value is the initial guess provided as input, and is never changed afterwards. Otherwise, \mygi\ begins scanning the \Teff-LEAS relationship. This can be performed in two modes: either each \Teff\ in the grid is tried (full scan mode), or only a few hundreds K around the initial estimate are tried (local scan mode). 
To do so, for each temperature probed \glog, \Vturb, \feh, and \alphafe\ are determined (steps between \ref{collapse} and \ref{alphameas} below), then the LEAS is determined (step \ref{getleas} below). The operation is repeated for each \Teff\ to be scanned. After the scan is completed, \mygi\ begins to refine its temperature estimate by fitting the \Teff-LEAS relationship with a 2nd order polynomial and estimating the zero-LEAS \Teff\ value. Each new attempt is added to the sample, the fit repeated and the new zero determined. Search stops when step-to-step \Teff\ change is less than a threshold parameter value (typically, 50K). 
\item \label{collapse} At the beginning of the actual line measurement stage, the synthetic grid is interpolated at the current guess values for \Teff, \glog, \Vturb, and \alphafe. As a consequence, it is now reduced to a set of synthetic spectra with equal atmosphere parameters but varying metallicity. 
\item \label{feimes} A set of \fei\ lines is measured: for each feature, the set of synthetic spectra at varying metallicity is compared against the observed spectrum and the best fit determined by $\chi^2$ minimization allowing three free parameters: metallicity, a small\footnote{Currently, one quarter of synthetic grid broadening. E.g. 1.75 kms$^{-1}$ for a grid broadened to 7 kms$^{-1}$.} radial velocity shift,
and a small deviation from the established continuum value, never to surpass a given fraction of the local S/N ratio. Local S/N is evaluated by fitting a 3rd order polynomial to all the S/N estimates in the relevant observed frame. {Also, EW is determined for both the observed and best fitting synthetic feature by direct integration under the pseudo-continuum. EWs will be used in some of the line rejection criteria, and in the search for the best microturbulence, see below.} {Measurement for each} feature is kept if a number of criteria are met: i) the fit probability should exceed a given threshold, ii) the equivalent width of the observed and fitted synthetic line should not differ by more than a given threshold, iii) both the observed and the synthetic EWs should exceed a given value, iv) both EWs should exceed a certain number of times the EW of a noise-dominated line\footnote{The EW of a noise-dominated line is computed as the EW of a triangular line whose depth corresponds to the local $1 \sigma$ S/N, and whose FWHM is the same as either the grid instrumental broadening, or a user-provided value.} and v) EW should not exceed a user-defined maximum value EW$_{\mathrm{max}}$, to allow avoiding using heavily saturated lines. If any of the above tests fails, the feature is marked for rejection.
The same rejection criteria will be applied also to features for every other elements later on in the analysis.
\item \label{feisiclip} Once all the assigned \fei\ features have been measured, {the average of their abundance} is computed, and $\sigma$-clipping performed (every feature deviating more than $n\sigma$ from the average is rejected, $n$ being set {as one of \mygi\ input parameters}). 
After the $\sigma$-clipping phase, a linear fit is performed in the EW-A(Fe) plane to determine microturbulence. Step \ref{feimes} and \ref{feisiclip} are performed at least four times: the first time measuring \fei\ lines at the grid's lowest microturbulence value, the second time at the highest, the third at the zero of the linear fit of the first two, the fourth at the zero of the 2nd order fit of the first three. In a fashion similar to what described for \Teff\ in step \ref{tcontrol} every time a new \Vturb\ is evaluated, it is added to the sample and a 2nd order polynomial is fitted again to the whole set. {We do not seek the minimum  slope of the linear EW-A(Fe) relationship, but
when it is smaller than the threshold set in the parameter file, microturbulence is considered determined, as well as \fei\ abundance.}  
\item \label{feiimes} Step \ref{collapse} is repeated again at the established microturbulence, and  \feii\ lines are measured. Their  average, $\sigma$-clipped abundance is compared with the {average} \fei\ abundance. If their discrepancy exceed the threshold set in the parameter file, a new gravity is estimated from the size of said discrepancy and steps \ref{collapse}, \ref{feimes}, and \ref{feisiclip} are repeated until a new value of microturbulence and \fei\ abundance are found with the new gravity. \fei\ and \feii\ abundances are compared again, and the process is repeated until convergence is reached.
\item \label{alphameas} With the current estimates of microturbulence and gravity, lines are measured for all the $\alpha$ {element ions} chosen to estimate $\alpha$ enhancement {(we speak of ions because, for instance, one might choose to estimate $\alpha$ enhancement from \ion{Mg}{i} and \ion{Ti}{ii})}. {Average, $\sigma$-clipped abundances for each ion} are determined, and their respective [X/Fe] are averaged to estimate [$\alpha$/Fe]. Steps \ref{collapse} to \ref{feiimes} are repeated and all the parameters established again at the newly estimated $\alpha$ enhancement. The $\alpha$ enhancement itself is computed again and compared with the previous value. The process is iterated until the difference is below the threshold set in the parameter file.
\item \label{getleas} Now the LEAS is determined with the current parameters, and \mygi\ goes back to step \ref{tcontrol} if \Teff\ is being derived.
\item Now the full set of atmospheric parameters is established, and abundances {are measured for all the features of all the other ions given in the linelists but not yet measured, again
by $\chi^2$ fitting of the line profiles. When more than one featrure is available for a ion they are averaged, $\sigma$-clipped, and averaged again to determine the final value of the abundance of that ion}.
\item Output files are created,
and \mygi\ moves on to analyze the next star.
\end{enumerate}

{Summarizing the procedure is composed
of several ``blocks'' that have to be iterated: steps 3-9 can be viewed as the ``temperature block'',
steps 5-6 are the ``metallicity block'', step 6 is the ``microturbulence block'',
steps 5-7 are the ``gravity block'' and steps 5-8 are the ``alpha-enhancement block''.}

\begin{table}
\caption{Test of the ability of \mygi\ to correctly retrieve \ion{Ca}{i} abundances deviating from the assumed solar-scaled composition.}             
\label{CaOffset:table}      
\centering                  
{
\begin{tabular}{c c r l}        
\hline\hline                 
Input   & offset from & Retrieved & $\sigma$ \\
{[Ca/Fe]} & model value & [Ca/Fe]   &          \\
\hline
--0.5 & --0.9 & --0.52 & 0.05 \\
--0.2 & --0.6 & --0.23 & 0.05 \\
  0.1 & --0.3 &   0.07 & 0.04 \\
  0.4 &   0.0 &   0.39 & 0.04 \\
  0.7 &   0.3 &   0.70 & 0.04 \\
  1.0 &   0.6 &   1.00 & 0.04 \\
  1.3 &   0.9 &   1.29 & 0.05 \\
\hline                      
\end{tabular}
}
\end{table}

\subsection{Comments on the \mygi\ workflow}
{The method for fixing the microturbulence is essentially the same as in \citet{gala}.
The advantage of automatising a ``classical'' approach, rather than a global
$\chi^2$ minimisation, such as done in SME \citep{valenti96} or TGMET  \citep{katz98}
is  that in the classical approach the different
diagnostics (for \Teff, \glog, microturbulence, abundances)  
are kept separate, in a global $\chi^2$ minimisation they are
all considered together, and it is diffcult to break
degeneracies. In  the case information
other than the spectra can be used (e.g. distances), some
parameters can be easily and neatly  fixed in the classical
approach, less so in a global approach.
\mygi\ has the same advantage over other, non $\chi^2$ based, 
global methods
\citep{A00,BJ2000,recio06} that are indeed very fast, 
but cannot easily break degeneracies.
\citet{vds} do not use $\chi^2$ fitting but
minimise another quantity, that is similar to $\chi^2$ (see their
section 4.1), as a consequence they cannot use
the theorems of $\chi^2$ to estimate the goodness of fit.
}


\section{Specifics and limitations}
\label{specifics}
\subsection{Features vs. lines} 
\label{featvslines}
\mygi\ operates on spectral {\em features} rather than spectral lines. Technically, the fitter at its core compares an observed spectral range against a set of synthetic profiles of the same range varying in \feh\ (more on this in Sect. \ref{MetNotAbu}), and determines the \feh\ value for which the fit is best. As such, \mygi\ can fit blended features without problems, but, at the same time, it is unable to perform deblending. 

The first characteristic comes in handy e. g. when the user wants to derive an overall metallicity from low resolution spectra, because complex blends can be used as metallicity indicators. Another example is the case in which important lines (e. g. the only line available for an element one wants to measure) are blended with some other feature. And of course, \mygi\ has no problem fitting lines affected by Hyper Fine Splitting (HFS), provided HFS has been included when the synthetic grid has been computed.  {Naturally}, the quality of the fit of a blend will depend on how well the atomic data of all the relevant features are known. It is also important to keep in mind that element-to-element ratios are kept fixed to the ones set in the grid (with the relevant exception of \alphafe), which could skew the result of fitting a blend if the two elements involved are not present in the star's atmosphere with a ratio similar to the one assumed in the grid. 

The inability to perform deblending is relevant in any situation where the EW of a line is important, the most obvious case being \Vturb\ determination, which uses the EW of \fei\ lines. \mygi\ determines two EWs for each feature it measures: one for the observed spectrum feature, and one for the best-fitting synthetic. Both are computed by direct integration under the local continuum value, and their difference is among the criteria used to reject a fit (see Sect. \ref{workflow}, point \ref{feimes}). For \fei\ features, EW is then used to estimate \Vturb. However, if a feature is a blend, its total EW will be too large in comparison to the associated abundance, skewing the \Vturb\ fit. For this reason, the user is allowed to decide which \fei\ features to use for \Vturb\ estimation, and must restrain to use only features corresponding to {\em bona fide} isolated \fei\ lines.

\begin{table*}
\caption{Determined abundances for \object{Sun}, \object{Arcturus}, and \object{HD 126681}.} 
\label{stars:abutable1}
\centering
\begin{tabular}{l r r l  r r l  r r l}
\hline \hline
 & \multicolumn{3}{l}{{\bf \object{Sun}:} \Teff=5794, \glog=4.55, } & \multicolumn{3}{l}{{\bf \object{Arcturus}:} \Teff=4293 \glog=1.84 } & \multicolumn{3}{l}{{\bf \object{HD 126681}} \Teff=5530, \glog=4.41,}\\ 
 & \multicolumn{3}{l}{\Vturb=1.07, \alphafe=0.05}                   & \multicolumn{3}{l}{\Vturb=1.44, \alphafe=0.19}                      & \multicolumn{3}{l}{\Vturb=0.50, \alphafe=0.38}           \\
\\
    & Number of & \feh\ or                & $\sigma$ & Number of & \feh\ or                & $\sigma$ & Number of & \feh\ or                & $\sigma$\\
    & features  & [X/Fe]$^{\mathrm{(a)}}$ &          & features  & [X/Fe]$^{\mathrm{(a)}}$ &          & features  & [X/Fe]$^{\mathrm{(a)}}$ &         \\
\hline \\
\fei         & 84 & -0.06 & 0.059 & 59 & -0.44 & 0.073 & 116 & -1.24 & 0.087 \\
\feii        & 14 & -0.05 & 0.094 & 10 & -0.44 & 0.083 & 14  & -1.23 & 0.069 \\
\\
\ion{Na}{i}  & 3  &  0.06 & 0.069 & 1  & 0.01  & --    & 4   & -0.13 & 0.140 \\
\ion{Mg}{i}  & -- & --    & --    & -- & --    & --    & 1   &  0.60 & --    \\
\ion{Al}{i}  & 3  & -0.04 & 0.073 & 2  & 0.13  & 0.177 & 1   &  0.18 & --    \\
\ion{Si}{i}  & 2  &  0.20 & 0.12  & 8  & 0.33  & 0.123 & 6   &  0.35 & 0.096 \\
\ion{Si}{ii} & -- & --    & --    & 2  & 0.34  & 0.239 & 1   &  0.57 & --    \\
\ion{Ca}{i}  & 8  &  0.02 & 0.074 & 1  & 0.05  & --    & 10  &  0.40 & 0.093 \\
\ion{Ti}{i}  & 23 &  0.04 & 0.086 & 4  & 0.20  & 0.100 & 22  &  0.31 & 0.120 \\
\ion{Ti}{ii} & 9  &  0.11 & 0.222 & 2  & 0.23  & 0.157 & 10  &  0.32 & 0.250 \\
\ion{Ni}{i}  & 18 &  0.02 & 0.153 & 6  & 0.02  & 0.126 & 15  & -0.03 & 0.172 \\
\hline
\end{tabular}
\tablefoot{
\tablefoottext{a}{[X/\fei] for neutral species, [X/\feii] for ionized. $\sigma$ is line-to-line scatter for Fe, propagated with corresponding $\sigma_{\mathrm{[Fe/H]}}$ for [X/Fe].}
}
\end{table*}

In a similar fashion, the user will indicate which \fei\ lines to employ for \Teff\ estimation. In a blend of two \fei\ features, for instance, one might not meaningfully associate one single lower energy value. Also, it is customary to refrain from using low excitation lines in \Teff\ estimation, given their general tendency to be prone to stronger departures from local thermodynamical equilibrium (LTE). In diagnostic plots for the test stars (Sect. \ref{stars}, Figs. \ref{diag:sun} trough \ref{diag:HD26}) lines used or rejected in the \Teff\ and \Vturb\ fitting are clearly indicated.

\begin{figure}
\centering
\includegraphics[width=8cm]{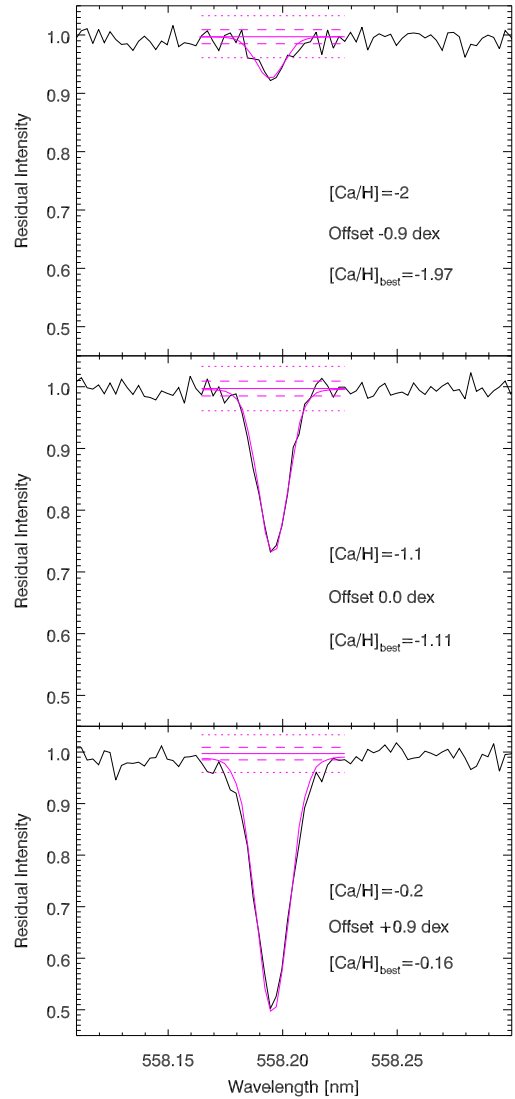}
\caption{\ion{Ca}{i} 559.19 nm line in the case of the two most extreme offsets described in Sect. \ref{MetNotAbu} and in the no-offset case, with their respective best fitting synthetic value (magenta profiles). The magenta horizontal lines represent local continuum, one-$\sigma$, and three-$\sigma$ of the local simulated noise. [Ca/H] is the input value, [Ca/H]$_{\mathrm{best}}$ the best fit value for the line.}
\label{CaOffset:figure}
\end{figure}

\begin{figure}
\centering
\includegraphics[width=8cm]{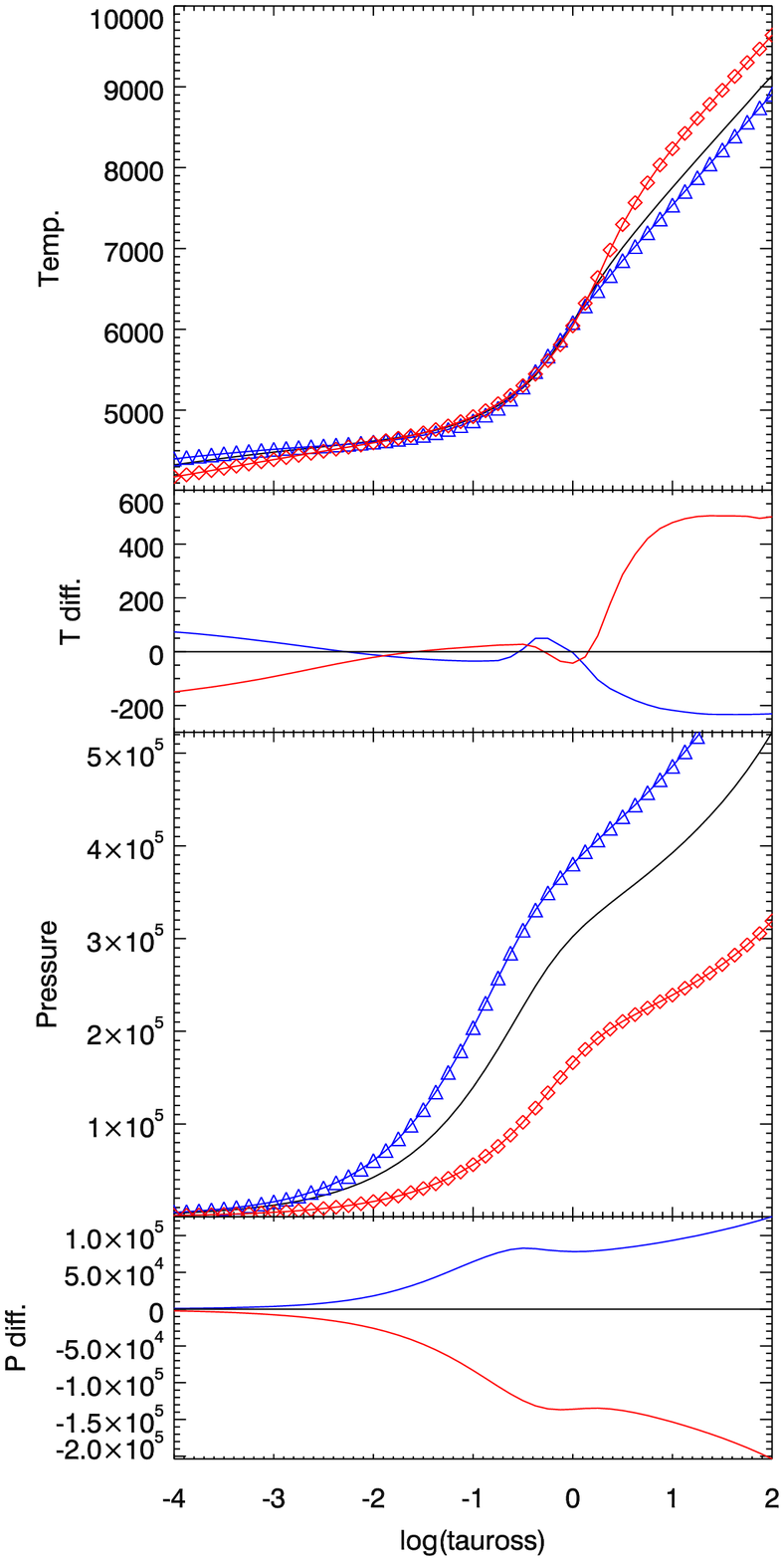}
\caption{Comparison of Atlas 12 atmosphere models with \Teff=5600 \glog=4.5, \Vturb=1.0, \alphafe=+0.4 and \feh=$-2.5$ (blue line and triangles), \feh=$-1.5$ (black line), and \feh=$-0.5$ (red line and diamonds).Top to bottom: Temperature for the three models, Temperature difference (red line, \feh=$-0.5$ -- \feh=$-1.5$; blue line, \feh=$-2.5$ -- \feh=$-1.5$), pressure in g cm$^{-2}$, pressure difference, all plotted vs. \tauross.}
\label{CaOffset:models}
\end{figure}

\subsection{Metallicity vs. abundance in line fitting}
\label{MetNotAbu}
\mygi\ computes abundances for lines of all elements by means of a grid which has only two degrees of freedom in chemical composition: metallicity and $\alpha$ enhancement. In fact, every line is fitted against metallicity only. Since all abundances scale the same way with metallicity in the grid (with the exception of $\alpha$ enhancement) \mygi\ interprets the result of the fit as due only to the change in the abundance of the specific ion producing the feature. By doing so, \mygi\ assumes that {\em varying [Fe/H] but keeping [X/Fe] constant produces the same effect on the line profile than varying [X/Fe] while keeping [Fe/H] constant}, or, in other words, it neglects the effect on the atmospheric structure of varying the overall star metallicity. This allows \mygi\ to drastically reduce the number of degrees of freedom in the grid, while still being able to measure abundances for an arbitrary number of ions. Otherwise, the grid should grow one dimension for every element which can be measured. This would either impose to use grids with a quite limited parameter span (which is undesirable when searching for optimal atmospheric parameters), or to use much larger grids, which are expensive to calculate, read in, and process inside \mygi. Moreover, with memory requirements being the bottleneck in running the code (see Sect. \ref{performa}), such very large grids would rapidly become unwieldy. 

The \mygi\ approximation thus remains valid as long as the measured abundance does not depart much from the grid solar-scaled composition, since this implies that the synthetic line profile is computed on the basis of an atmosphere which is not much different from the one providing the overall best parameter fit. To provide an indicative estimate of how ``safe'' the approach is, we have computed a series of synthetic spectra around parameters typical of a metal poor dwarf (\Teff=5600 K, \glog=4.5, \Vturb=1.0, \feh=--1.5, \alphafe=+0.4) and changing the Ca abundance to offsets of $\pm$ 0.3, 0.6, and 0.9 dex (since in the starting model [Ca/Fe]=+0.4, this corresponds to [Ca/Fe]=-0.5 to +1.3, and [Ca/H]=--2.0 to --0.2). The spectra were noise-injected to S/N=80 and fed to \mygi. First, the spectrum without Ca offset was analyzed with full parameter determination to determine how close \mygi\ would go to the input values, returning \Teff=5580 K, \glog=4.41, \Vturb=0.88, \feh=$-1.49$, \alphafe=0.39, values well in line with the typical offset to be expected as a result of noise injection (see \ref{montecarlo:sn}). Then parameters were locked to the input values, and \mygi\ was used to derive abundances only, to see whether the Ca offsets would be retrieved properly despite the Abundance-vs.-Metallicity approximation. The result of the test are presented in Table \ref{CaOffset:table} and, for a specific \ion{Ca}{i} line, in Fig. \ref{CaOffset:figure}. For the specific case, the approximation applied appears to produce no detectable effect on the ability of \mygi\ to properly measure abundances deviating for the assumed solar-scaled composition.

\begin{figure}
\centering
\includegraphics[width=8cm]{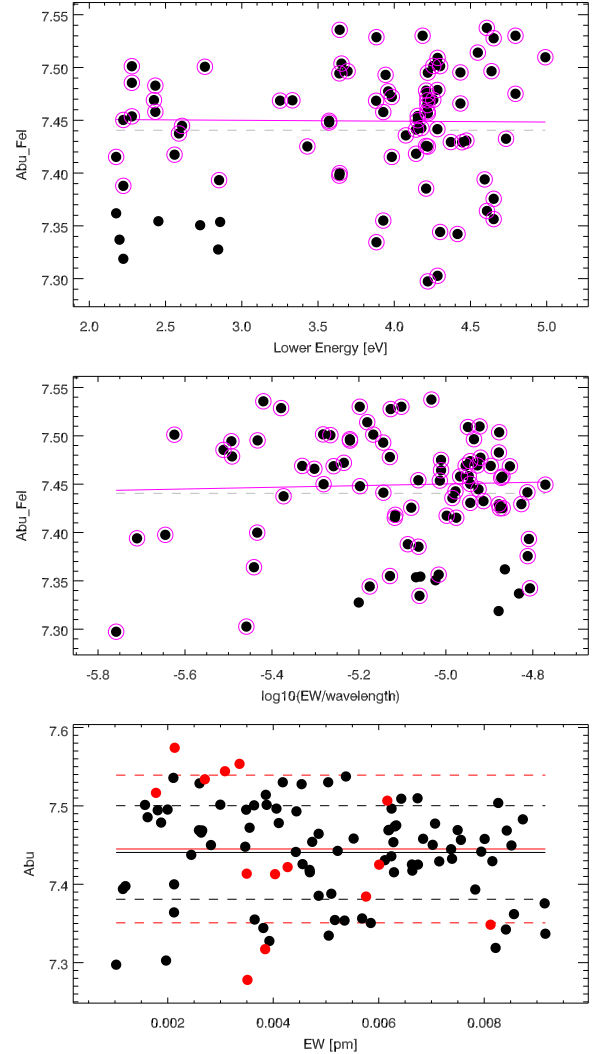}
\caption{Diagnostic plots for the determination of atmospheric parameters for the Sun. Upper panel: \fei\ lines abundances vs. their lower energy and linear fit for the determination of \Teff. Gray dashed line, average \fei\ abundance. The magenta line is the linear fit, magenta-circled points are the ones considered in the fit. LEAS is $-8.0~10^{-4}$ per eV. Center panel, \fei\ lines abundances vs. reduced EW (i. e. $\log_{10}\left( EW / \lambda \right)$ ) for the determination of \Vturb. Gray dashed line, average \fei\ abundance. Magenta symbols and line as in the top panel. Linear fit slope is $8.7~10^{-3}$. Bottom panel, \fei\ (black) and \feii\ (red) lines abundances for the check of gravity through ionization equilibrium. Continuous line represent averages, dashed lines $1\sigma$ intervals. A(\fei)$=7.44~\pm 0.06$, A(\feii)$=7.45~\pm 0.09$.}
\label{diag:sun}
\end{figure}

We want to stress that the presented case is not necessarily representative for any possible line being analyzed. For instance, if the line being fitted is significantly blended, in the synthetic grid being fitted against the observed both the ``main'' feature and the blending ones will vary, which would lead to skewed resulting abundances for any element deviating significantly form solar-scaled abundance. Also, the degree of sensitivity to the \mygi\ approximation depends on the physics of line formation for the specific feature. In Fig. \ref{CaOffset:models} we show the comparison of three atmospheric models from our grid very close to the ones of interest in this comparison. The ``central'' model is the one used to compute all the simulated observed spectra considered in this section (\Teff=5600 K, \glog=4.5, \Vturb=1.0, \feh=$-1.5$, \alphafe=+0.4). The other two differ in metallicity only, being offset by 1 dex upards and downwards, thus being representative of the $\pm$ 0.9 dex variation in metallicity of the models used in the fitting grid. In other terms, while all the synthetic profiles being fitted to our simulated observation, in the ideal case, should be drawn from the ``central'' model by only varying Ca abundance, they are instead drawn by models which differ in metallicity as much as the ones plotted in Fig. \ref{CaOffset:models}.

\begin{figure}
\centering
\includegraphics[width=8cm]{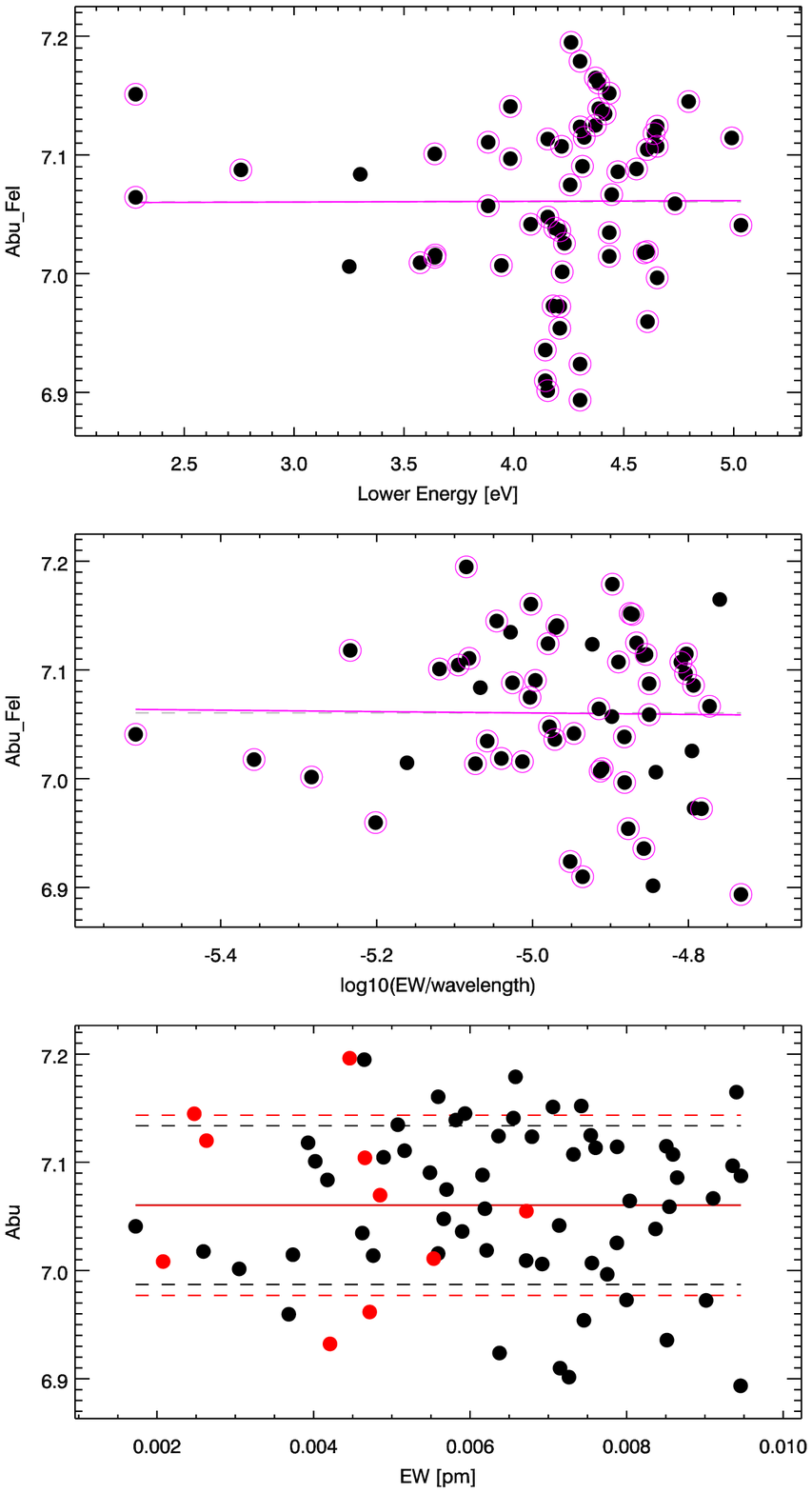}
\caption{Same as in Fig. \ref{diag:sun} for \object{Arcturus}. LEAS=0.0006 per eV. \fei\ abundance vs. reduced EW slope $-6.4~10^{-3}$. A(\fei)$=7.06~\pm 0.07$, A(\feii)$=7.06~\pm 0.08$}
\label{diag:arct}
\end{figure}

\ion{Ca}{i} lines as the ones measured in this example belong to a trace specie in such atmospheres, and are thus temperature sensitive but only weakly pressure sensitive. At the typical formation depths for weak lines ($\log ($\tauross$)$ roughly between --2 and 0) the three models are quite close in temperature. Pressure on the other hand is significantly different, but these lines are very weakly sensitive to it. Majority species lines (e.g. ionized element lines), or damped lines which have stronger pressure sensitivity, are likely to display stronger departures in this situation. As such, we suggest that users verify the abundances derived by \mygi\ for species which depart heavily from grid solar-scaled values. However, since significant differences might arise only for species strongly departing from solar abundance ratios, the {\em detection} of abundance anomalies through \mygi\ is to be considered robust, and only the exact amount of the abundance departure might be in need of verification.

\begin{figure}
\centering
\includegraphics[width=8cm]{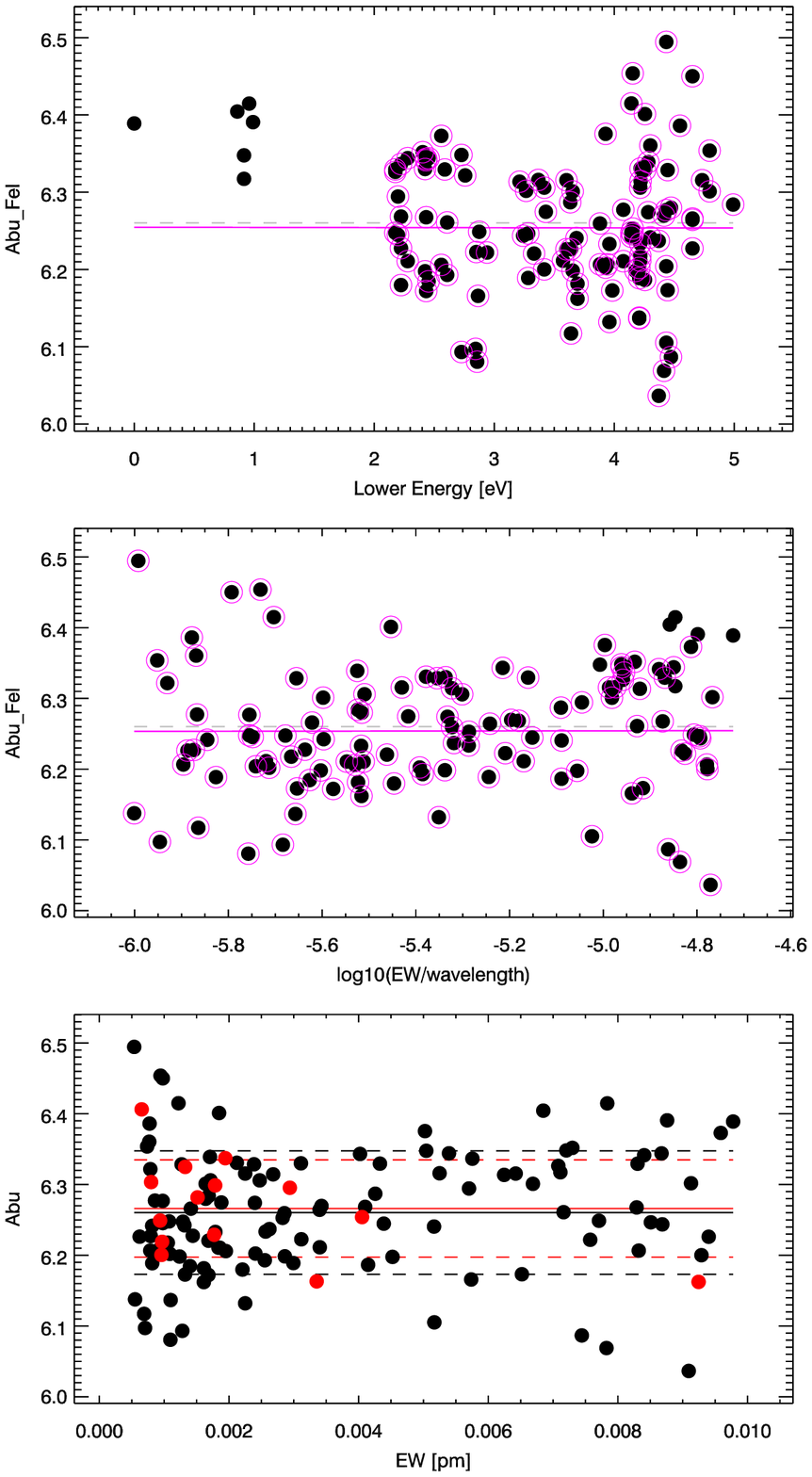}
\caption{Same as in Fig. \ref{diag:sun} for \object{HD 126681}. LEAS=-0.0002 per eV. \fei\ abundance vs. reduced EW slope $8.0~10^{-4}$. A(\fei)$=6.26~\pm 0.09$, A(\feii)$=6.27~\pm 0.07$}
\label{diag:HD12}
\end{figure}


\section{Tests on reference stars}
\label{stars}

As an assessment of \mygi\ performance, we present in this section parameter determination and chemical analysis for a few representative stars. To reproduce typical data characteristics, very high S/N, high resolution spectra of five well studied stars have been noise-injected at typical ``real world'' values and analyzed using synthetic grids and feature lists appropriate for the four main broad spectra type: metal rich dwarf / subgiant stars (the \object{Sun}, see Sect. \ref{stars:sun}), metal poor dwarf / subgiants (\object{HD 126681}, Sect. \ref{stars:12}, and \object{HD 140283}, Sect. \ref{stars:14}), metal rich giants (\object{Arcturus}, Sect. \ref{stars:arct}), and metal poor giants (\object{HD 26297}, Sect. \ref{stars:26}). Given the coverage provided by the available synthetic grid, the analysis has been performed using ranges roughly corresponding to UVES red 580nm setting, i.e. two frames covering 480 to 580nm, and 580 to 680nm.
For each star, Table \ref{stars:abutable1} and \ref{stars:abutable2} report the final determined parameters, plus detailed abundances for a few chemical species, together with the number of lines used and line-to-line scatter, when more than one line was measured. 

\begin{figure}
\centering
\includegraphics[width=8cm]{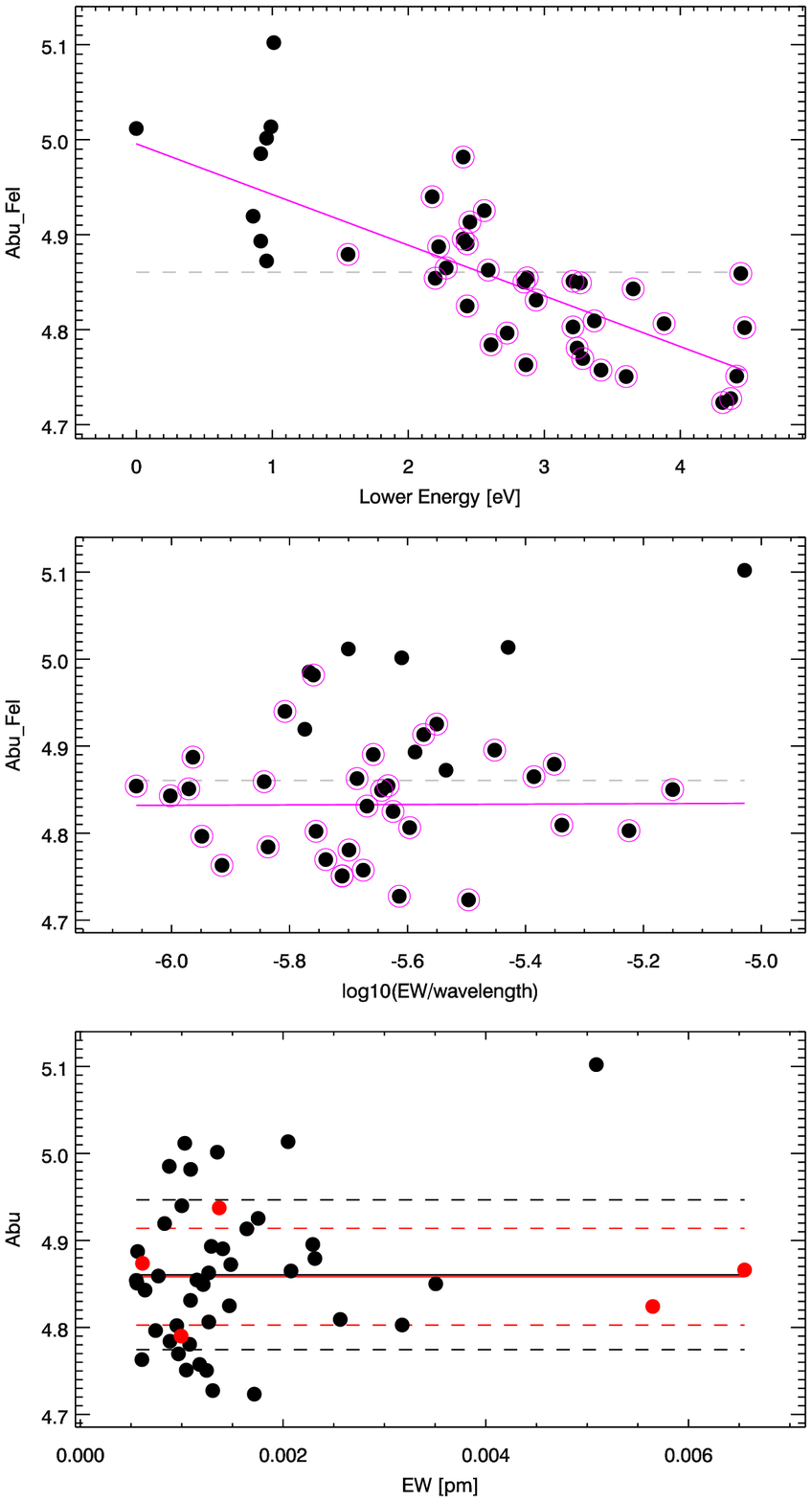}
\caption{Same as in Fig. \ref{diag:sun} for \object{HD 140283} with temperature locked at \Teff=5777 K. LEAS=-0.0534 per eV. \fei\ abundance vs. reduced EW slope $2.3~10^{-3}$. A(\fei)$=4.86~\pm 0.09$, A(\feii)$=4.86~\pm 0.06$}
\label{diag:HD14l}
\end{figure}

\subsection{The \object{Sun}}
\label{stars:sun}

Diagnostic plots for the \object{Sun} are shown in Fig. \ref{diag:sun}, abundances listed in Table 
\ref{stars:abutable1}.
A high S/N ($>$100) spectrum of \object{Ceres} was acquired (on 18/01/2008, UT 18:00:56, 900 s exposure) at SOPHIE@OHP \citep[][]{bouchy06,perruchot08} in High Efficiency mode (resolution of 7.4 km s$^{-1}$), and was degraded to UVES-like sampling ($\sim$ 0.0025 nm per pixel) and noise-injected to S/N=50. 
Derived parameters (\Teff=5794, \glog=4.55, \Vturb=1.07 km s$^{-1}$, \alphafe=0.05, \feh= --0.06 $\pm$0.059 over 84 \fei\ features) are in excellent agreement with the Sun's  effective 
temperature and gravity
(\Teff=5777 K, \glog=4.44, \citealt{cox}).
While the ``canonical'' value of the solar microturbulence for the
full disk flux spectrum is 1.35 km s$^{-1}$ \citep{holweger}
we note that an analysis conducted at lower resolution and based on 
an ATLAS 9 model derives 0.99  km s$^{-1}$ \citep{melendez}.

The difference from the established values are well within the dispersion of random S/N=50 noise injections, as presented below in \ref{montecarlo:sn}. 

\subsection{\object{Arcturus}}
\label{stars:arct}

The UVES POP \citep[][]{bagnulo03} spectrum for \object{Arcturus} derived with the Red UVES arm and 580nm setting was convolved with a gaussian to produce a spectrum with final resolution of 7.0 km s$^{-1}$, then noise-injected at S/N=50. 
Diagnostic plots are shown in Fig. \ref{diag:arct}, abundances listed in Table \ref{stars:abutable1}. The derived parameters (\Teff=4293, \glog=1.84, \Vturb=1.44 km s$^{-1}$, \alphafe=0.19, \feh=--0.44 $\pm$0.073 over 59 \fei\ features) are in excellent agreement with the values presented in \citet{koch08} (\Teff=4290, \glog=1.64, \Vturb=1.54, \feh=--0.49).

\begin{figure}
\centering
\includegraphics[width=8cm]{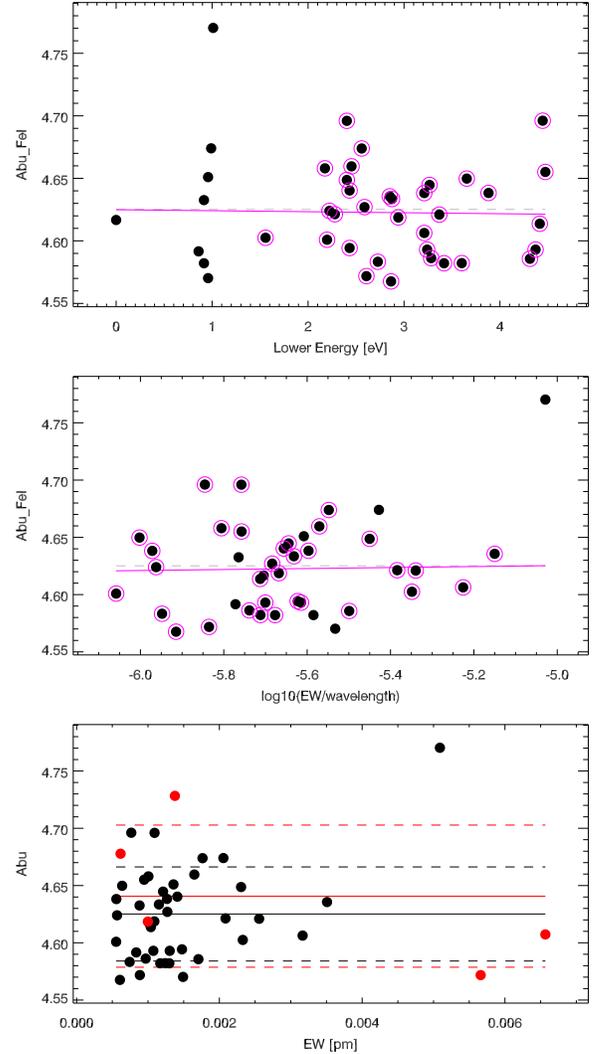}
\caption{Same as in Fig. \ref{diag:sun} for \object{HD 140283} with temperature iterated. LEAS=--0.0008 per eV. \fei\ abundance vs. reduced EW slope $4.3~10^{-3}$. A(\fei)$=4.63~\pm 0.04$, A(\feii)$=4.64~\pm 0.06$}
\label{diag:HD14f}
\end{figure}

\subsection{\object{HD 126681}}
\label{stars:12}

\object{HD 126681} is a good example of a moderately metal-poor dwarf star. A ``traditional'' fully spectroscopic parameter determination for this star, based on high resolution and S/N HARPS spectra is presented in \citet{sousa11}. It is the very type of analysis \mygi\ aims to replicate, and leads to values of \Teff=5561 $\pm$68 K, \glog=4.71 $\pm$0.1, \Vturb=0.71 km s$^{-1}$, \feh=--1.14$\pm$0.06. Similarly to what we did for \object{Arcturus}, we have employed the UVES POP spectrum, degraded to a resolution of 7.4 km s$^{-1}$, and noise-injected to S/N=80. 
Diagnostic plots for the \object{HD 126681} are shown in Fig. \ref{diag:HD12}, abundances listed in Table \ref{stars:abutable1}. The derived values (\Teff=5530, \glog=4.41, \Vturb=0.50, \alphafe=0.38, \feh=--1.24 $\pm$0.087 over 116 features) are in very good agreement with the above stated values. The \feh\ discrepancy is a bit higher than expected, our metallicity being lower by 0.10 dex. It is worth noting how a change upwards of \fei\ abundance alone would lead ionization equilibrium to settle to a somewhat higher gravity, as found by \citet{sousa11}. We have not further investigated this slight discrepancy, which can stem from a number of differences, including differences in \fei\ line data, or the fact that \citet{sousa11} use a model grid with a different set of opacities and the
overshooting  option switched on. 
The difference between overshooting and non-overshooting ATLAS models
can account for a 0.1\, dex difference.

\subsection{HD 140283}
\label{stars:14}

\object{HD 140283} is the prototypical
metal-poor star \citep{chamberlain51},
very often employed as a reference for stars in this metallicity range \citep[recently,][among others]{hosford09,hosford10,casagrande10,bergemann12}. Recently its distance has been estimated  through HST observations \citep{bond13}, allowing to estimate an age of 14.46 $\pm 0.8$ Gyr. 
\citet{hosford09} noticed how attempting to constrain ionization and excitation equilibrium together led to a relatively low temperature (5573 K) and consequently to a low gravity incompatible with the known Hipparcos-based luminosity. 
The InfraRed Flux Method (IRFM) based temperature 
(5691\, K {Alonso}, 5755\,K \citealt{JGHB}, 5777\,K \citealt{casagrande10}) delivers a more satisfactory ionization equilibrium when gravity is derived from the Hipparcos luminosity, but at the expense of inducing a sizeable LEAS \citep[see Tables 2, 3 and figure 11 in][]{bergemann12}. 
According to \citet{hosford10}, treating the line transfer
for \fei\ lines in NLTE solves this discrepancy and provides an 
excitation temperature of 5838\,K, in reasonable agreement with the
IRFM temperatures. \citet{bergemann12} disagree with this finding. 
What is relevant in the present context is that solving for both excitation
and ionization equilibria of iron in LTE for HD\,140283 leads to low temperature 
and gravity and on this issue \citet{hosford09} and \citet{bergemann12}
agree.

We employed the UVES POP red arm 580 nm spectra, without adding instrumental broadening or injecting noise: given the high resolution, 
stellar rotation and macroturbulence might require a broadening
of the grid above what necessary to take into 
account the instrumental resolution. We adjusted the grid broadening
until we obtained satisfactory fits on several isolated lines. The final grid broadening employed was 5.5 km s$^{-1}$.
We ran \mygi\ on this star twice, once leaving the code free to constrain all the parameters, and once locking \Teff\ to 5777 K. Diagnostic plots for the \object{HD 140283} are shown in Fig. \ref{diag:HD14l} and \ref{diag:HD14f}, abundances listed in Table \ref{stars:abutable2}.  
When leaving \mygi\ free to set \Teff, it falls to 5506 K, 
resonably close to the excitation
temperature found by \citet{hosford09} 
driving also down \feh\ to --2.87, and \glog\ to 2.86.  
When fixing \Teff\ at 5777 K, instead, the derived gravity (3.32) is in reasonable
agreement with the Hipparcos value (3.73), and \Vturb\ is also very close to the value derived in \citet{bergemann12} in the LTE case (1.22 km s$^{-1}$ vs. 1.27). However, the metallicity we derive is lower by about 0.22 dex (\feh=--2.64 vs. --2.42).

\subsection{HD 26297}
\label{stars:26}

\object{HD 26297} is a moderately metal-poor giant star. A spectroscopic abundance analysis similar to the one \mygi\ performs is presented in \citet{fulbright00} (\Teff=4500 K, \glog=1.2, \feh=--1.72), while \citet{gratton00} obtains similar values (\Teff=4450 K, \glog=1.18, \feh=--1.68, \Vturb=1.62 km s$^{-1}$) but deriving temperature and gravity from photometric calibrations (Fe ionization equilibrium is very well satisfied for this star). A more recent analysis, presented in \citet{prugniel11}, uses instead a $\chi^2$ minimization algorithm to fit the target spectrum agains the empirical ELODIE library \citep[see][]{wu11}, and delivers \Teff=4479 K, \glog=1.05, \feh=--1.78.

\begin{figure}
\centering
\includegraphics[width=8cm]{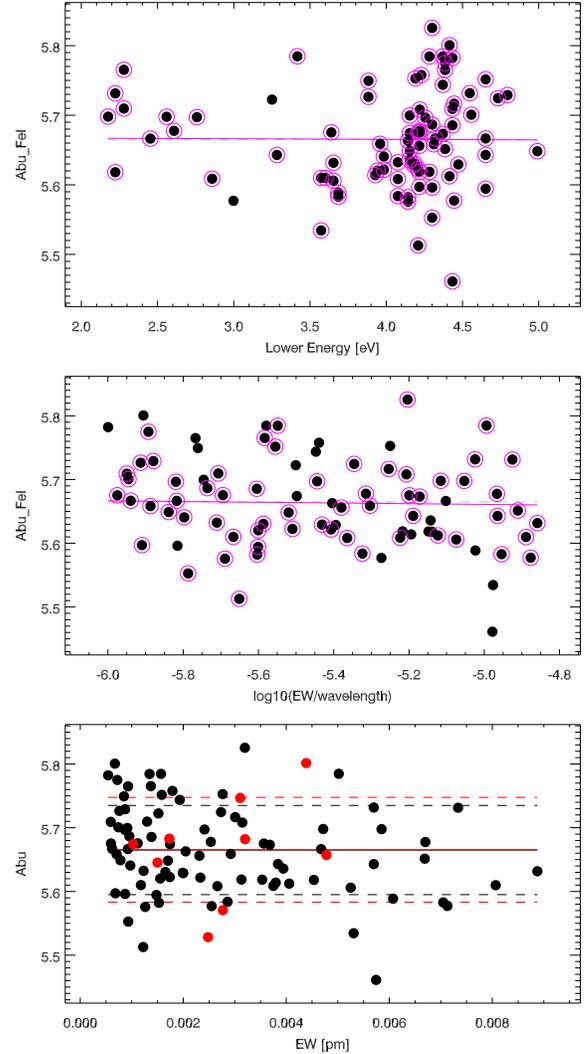}
\caption{Same as in Fig. \ref{diag:sun} for \object{HD 26297}. LEAS=0.0009 per eV. \fei\ abundance vs. reduced EW slope $-5.6~10^{-3}$. A(\fei)$=5.67~\pm 0.07$, A(\feii)$=5.67~\pm 0.08$}
\label{diag:HD26}
\end{figure}

We downloaded archival UVES red 580 spectra for \object{HD 26297} (taken on 20/11/2004, 01:39:42 UT, prog. id 074.B-0639). The data were taken with slit width of 0.9'' and have a good S/N ($>$100). No additional noise was injected. They were analyzed by broadening the synthetic grid to 7.5 km s$^{-1}$.
Diagnostic plots are shown in Fig. \ref{diag:HD26}, abundances listed in Table \ref{stars:abutable2}. We derive \Teff=4458, \glog=1.06, \Vturb=1.76, \alphafe=0.36, \feh=$-1.83\pm$0.07 over 83 \fei\ lines, in excellent agreement with all the cited sources.

\section{Montecarlo simulations}
\label{montecarlo}

\begin{table*}
\caption{Determined abundances for \object{HD 140283} with fixed and free \Teff, and \object{HD 26297}.} 
\label{stars:abutable2}
\centering
\begin{tabular}{l r r l  r r l  r r l}
\hline \hline
 & \multicolumn{3}{l}{{\bf \object{HD 140283}$^{\mathrm{(a)}}$:} \Teff=5777, \glog=3.32, } & \multicolumn{3}{l}{{\bf \object{HD 140283}$^{\mathrm{(b)}}$:} \Teff=5506 \glog=2.85 } & \multicolumn{3}{l}{{\bf \object{HD 26297}} \Teff=4458, \glog=1.06,}\\ 
 & \multicolumn{3}{l}{\Vturb=1.22, \alphafe=0.26}                                          & \multicolumn{3}{l}{\Vturb=1.33, \alphafe=0.35}                                        & \multicolumn{3}{l}{\Vturb=1.76, \alphafe=0.36}           \\
\\
    & Number of & \feh\ or                & $\sigma$ & Number of & \feh\ or                & $\sigma$ & Number of & \feh\ or                & $\sigma$\\
    & features  & [X/Fe]$^{\mathrm{(c)}}$ &          & features  & [X/Fe]$^{\mathrm{(c)}}$ &          & features  & [X/Fe]$^{\mathrm{(c)}}$ &         \\
\hline \\
\fei         & 41 & --2.64 & 0.086 & 41 & --2.87 & 0.041 & 85  & --1.83 & 0.070 \\
\feii        & 5  & --2.64 & 0.056 & 5  & --2.86 & 0.062 & 9   & --1.83 & 0.082 \\
\\
\ion{Na}{i}  & -- &  --    & --    & -- &  --    & --    & 3   & --0.41 & 0.102 \\
\ion{Mg}{i}  & 1  &   0.49 & --    & 1  &  0.57  & --    & 1   &   0.53 & --    \\
\ion{Al}{i}  & -- &  --    & --    & -- &  --    & --    & 1   & --0.06 & --    \\
\ion{Si}{i}  & -- &  --    & --    & -- &  --    & --    & 6   &   0.26 & 0.077 \\
\ion{Ca}{i}  & 6  &   0.26 & 0.093 & 6  &  0.35  & 0.044 & 9   &   0.28 & 0.075 \\
\ion{Ti}{i}  & 4  &   0.40 & 0.087 & 4  &  0.33  & 0.043 & 16  &   0.22 & 0.097 \\
\ion{Ti}{ii} & 4  &   0.28 & 0.182 & 4  &  0.32  & 0.261 & 5   &   0.23 & 0.356 \\
\ion{Ni}{i}  & 2  &   0.11 & 0.243 & 2  &  0.11  & 0.155 & 10  & --0.11 & 0.101 \\
\hline
\end{tabular}
\tablefoot{
\tablefoottext{a}{\Teff\ kept fixed at 5777 K;}
\tablefoottext{b}{\Teff\ determined by \mygi;}
\tablefoottext{c}{[X/\fei] for neutral species, [X/\feii] for ionized. $\sigma$ is line-to-line scatter for Fe, propagated with corresponding $\sigma_{\mathrm{[Fe/H]}}$ for [X/Fe].}
}
\end{table*}

One attractive possibility opened by fully automated abundance analysis codes such as \mygi\ is to comprehensively assess the error budget by means of Montecarlo simulations, since large number of test ``events'' can be processed through the code in very little time. 

   \begin{figure}
   \centering
   \includegraphics[width=7.5cm]{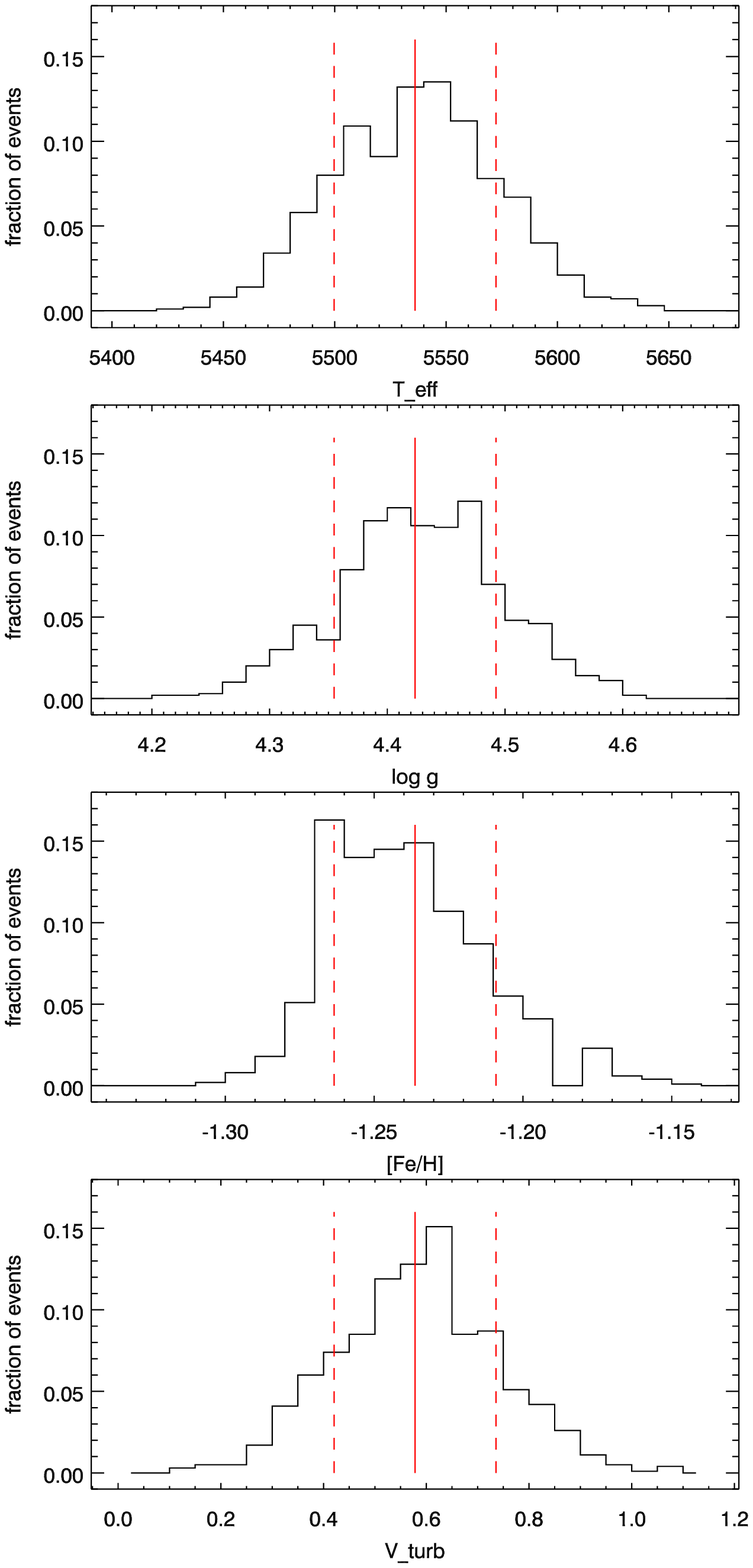}
      \caption{Histograms of \mygi\ parameter determination output in a Montecarlo simulation of 1000-events noise injection at S/N=80 on a spectrum of \object{HD 126681}. Top to bottom: determined \Teff\ (mean and $\sigma$, continuous and dashed red lines, respectively, all values in Table \ref{montecarlo:table}), \glog, \feh, \Vturb.}
         \label{histo:sn80}
   \end{figure}

{\subsection{Impact of S/N ratio: internal errors.}
\label{montecarlo:internal}
To assess the size of \mygi\ internal errors, as well as the extent to which the code is affected by spectrum noise, we have employed a synthetic spectrum extracted from the MPD grid at \Teff=5400 K, \glog=4.5, \Vturb=1.0 kms$^{-1}$, \feh=-1.0, \alphafe=+0.4, and have processed it to \mygi\ after broadening it to a resolution of 7.4 kms$^{-1}$, and resampling it to typical UVES pixel size. We have first processed it without injecting noise to ensure the absence of systematics, and subsequently produced 1000 independent Poisson noise realizations at S/N=80, 50, 20, for a total of 3000 noise realizations, which were fed to \mygi, which was set to derive \Teff, \glog, \Vturb, \feh, and \alphafe. The results of the noiseless test, as well as derived average values and $1 \sigma$ dispersions  for the noise-injection montecarlo tests are listed in Table \ref{montecarlo:table}.
}

{It is worth noticing how the errors displayed by \mygi\ in the synthetic test match closely the ones in the case of \object{HD126681}. One might have expected that feeding \mygi\ a spectrum of the same grid used for the analysis would ahve led to the code ``snapping'' on the right parameters, leading to lower-than-realistic errors. However, \mygi\ only compares the grid and the observed on a line-by-line basis, and parameters are all derived by indirect methods. Moreover, the initial temperature scan is purposefully performed at steps different from the grid step, so that comparison is always performed away from gridpoints, removing any ``grid snapping'' effect.}

\subsection{Impact of S/N ratio: {\object{HD126681}}}
\label{montecarlo:sn}

\begin{figure}
   \centering
   \includegraphics[width=7.5cm]{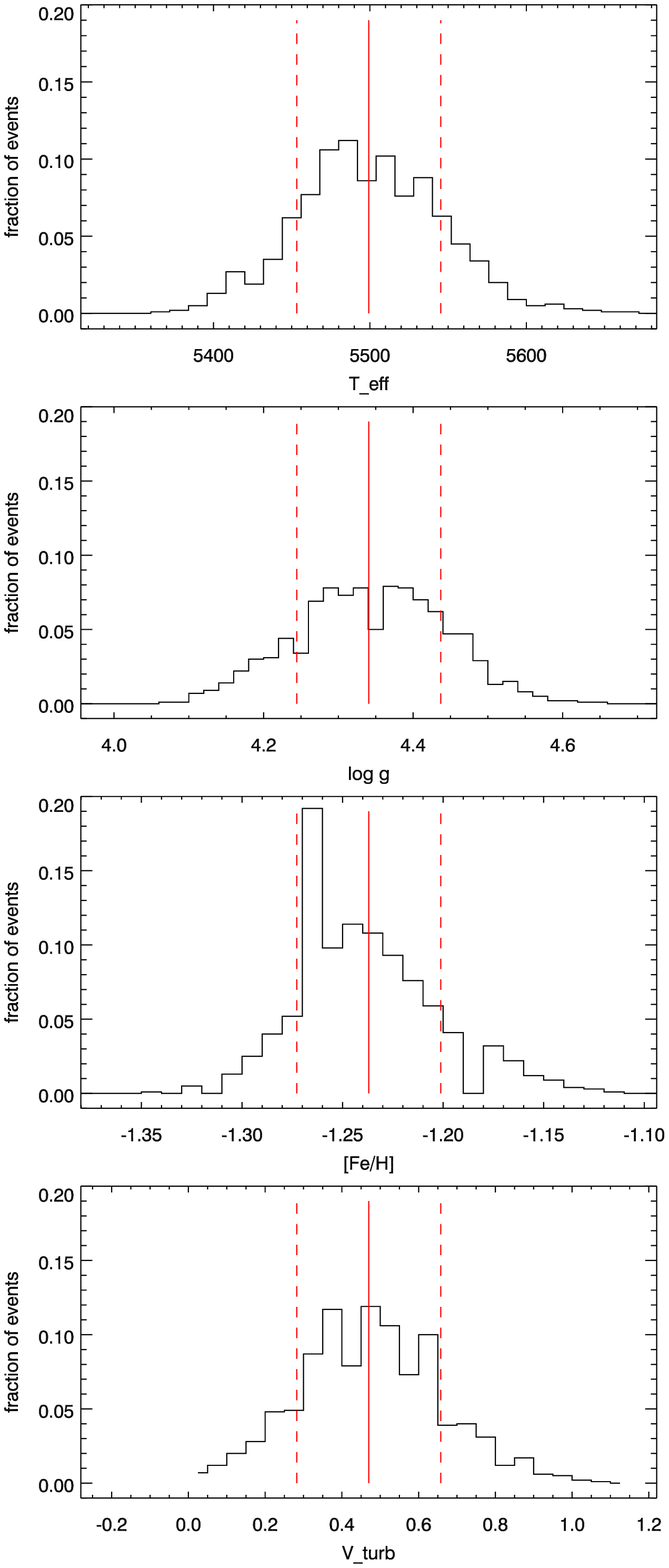}
      \caption{As in Fig. \ref{histo:sn80} but now with S/N=50.}
         \label{histo:sn50}
   \end{figure}

\begin{figure}
   \centering
   \includegraphics[width=7.5cm]{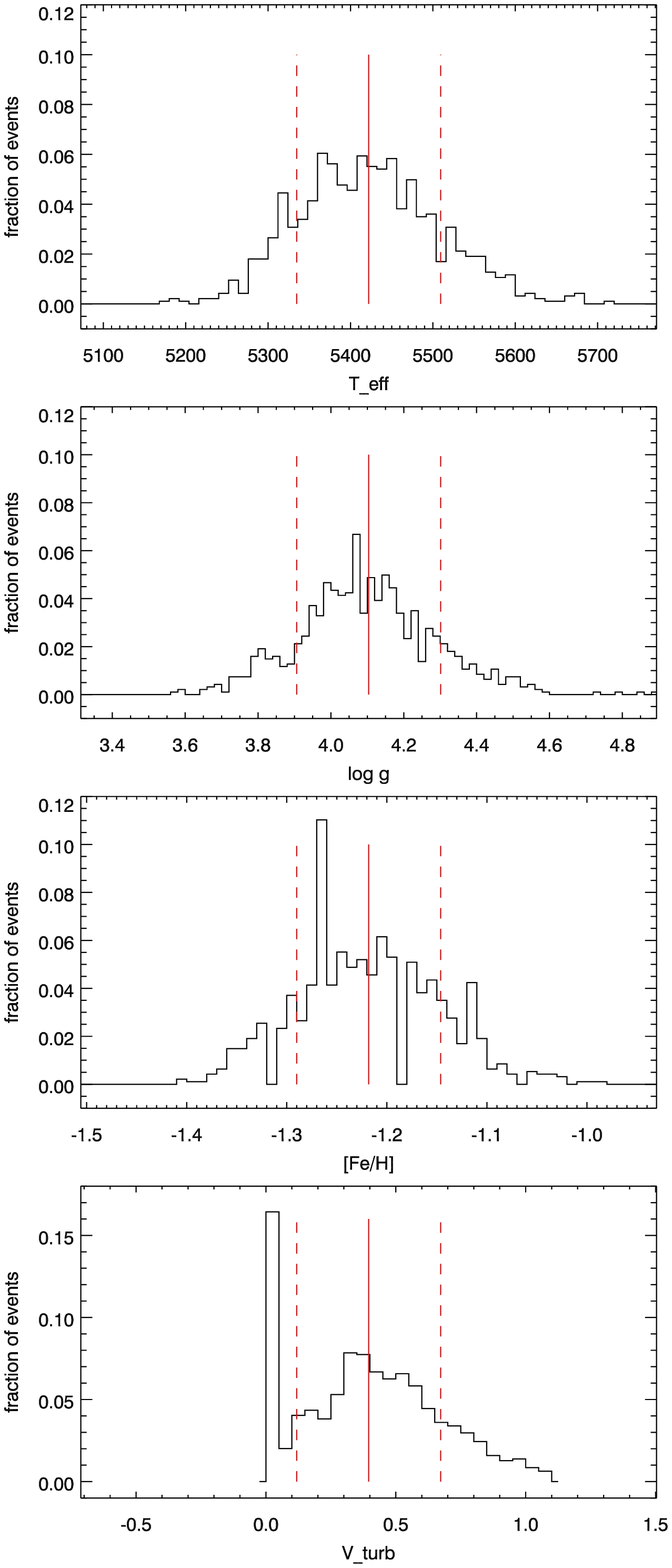}
      \caption{As in Fig. \ref{histo:sn80} but now with S/N=20.}
         \label{histo:sn20}
   \end{figure}

{To assess the impact of external error sources we have repeated the above described test on the very high S/N spectrum of \object{HD126681} described in \ref{stars:12}. Again, 1000 independent Poisson noise realization were produced for S/N=89, 50, 20 each. Derived average values and $1 \sigma$ dispersion sare listed in Table \ref{montecarlo:table}, the histograms of the parameters are presented in Fig. \ref{histo:sn80} through \ref{histo:sn20}.}

\mygi\ is not allowed to try indefinitely to bring stellar parameters to convergence. After a set number of ``cycles'' it stops, declares the star 
non-converging and passes to the next one. It is worth noticing here that, while S/N=80 and 50 Montecarlo tests had every single star reaching satisfactory convergence, in the S/N=20 case convergence was reached in 943 over 1000 realizations (a failure rate of 5.7\%). Unsurprisingly, at such low S/N the difficulty of measuring weak lines, needed especially for \Vturb\ determination, begins to take its toll. Moreover, 138 converging stars end up with \Vturb=0, a clearly visible peak in Fig. \ref{histo:sn20}. This is just the natural consequence of a \Vturb\ distribution centered at 0.39 km s$^{-1}$ with a $\sigma$ of 0.28: in a relevant number of stars the specific set of \fei\ lines used would push for a negative \Vturb, which \mygi\ forbids as unphysical. 
   
It is to be remarked that the presented test is a simplified one and the resulting uncertainties are possibly somewhat underestimated. In particular, the injection of Poisson noise is a somewhat ``optimistic'' way to simulate real observed spectrum noise degradation, since it lacks a number of realistic outstanding defects real, low S/N spectra present (cosmic ray hits, poor order tracing in low S/N cases, increased noise at order merging points...). Also, noise has been injected here to produce constant S/N through the range, which is not the case in practice, often with lower S/N in the bluer part of the range, where most of the atomic lines are concentrated.

\subsection{Impact of pre-determined parameters: photometric \Teff}
\label{montecarlo:phototemp}

Most abundance analysis works typically assess the impact of errors on atmospheric parameters on derived abundances by assuming said errors are independent. This is often performed by picking a representative star in the sample, and repeating the abundance determination by varying {\em one parameter at the time} by an amount considered to be the typical uncertainty on that parameter. This quick way of estimating parameter uncertainties is, however, conceptually flawed in the sense that atmospheric parameters correlate in a fashion that is, ultimately, dependent on the specific way they are derived. For instance, if gravity is determined from \fei-\feii\ ionization equilibrium, varying \Teff\ by 150 K will lead to a different gravity too: repeating the analysis with a different temperature but the same gravity will not be representative of which effect a 150 K \Teff\ error  would have in a ``real world'' case. 

\begin{figure}
   \centering
   \includegraphics[width=7.5cm]{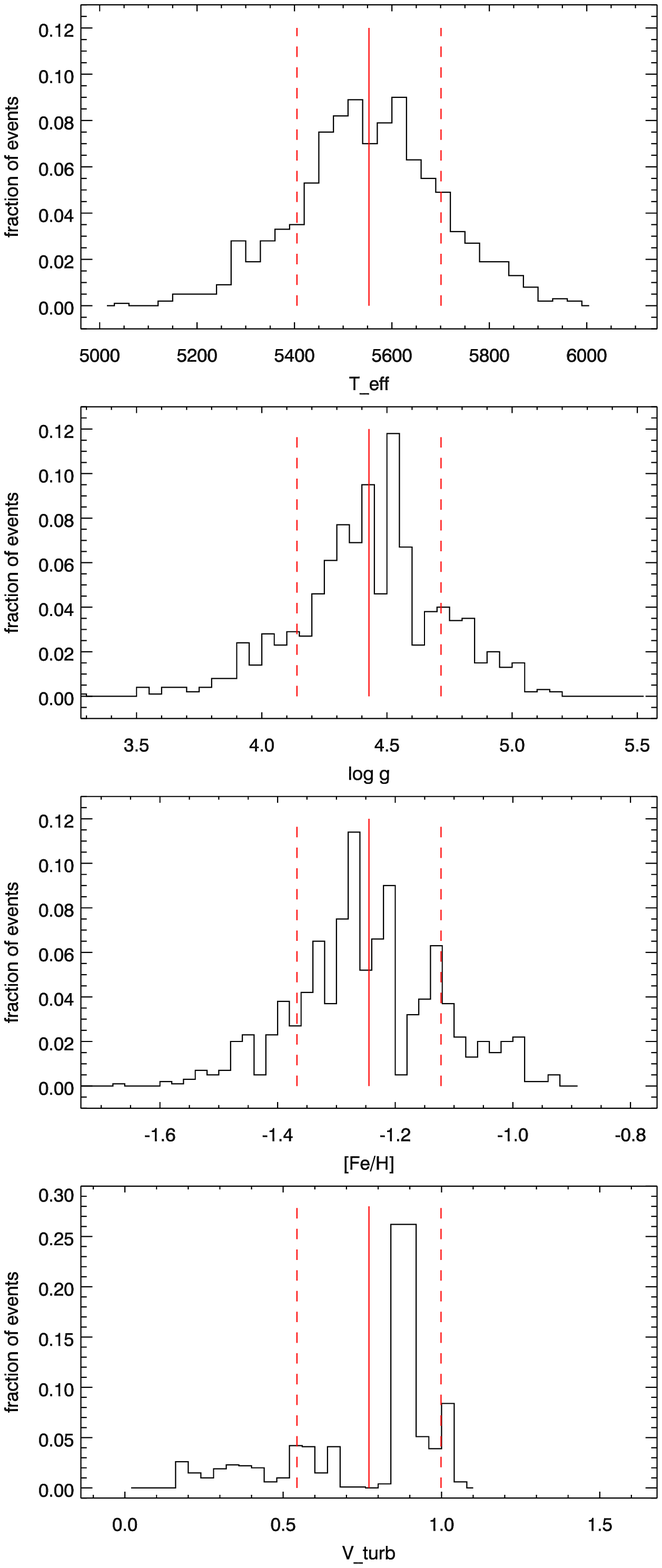}
      \caption{Histograms of \mygi\ parameter determination output in a 1000-events Montecarlo simulation with fixed \Teff, on a S/N=80 spectrum of \object{HD 126681}. Panels as in Fig. \ref{histo:sn80}.}
         \label{histo:phototemp}
   \end{figure}

A fairly common case in which such issue is of importance is the one in which effective temperatures are derived 
independently from the spectra, e.g. from photometry. In this case, errors in the other parameters will cascade from whatever error is made in \Teff, in a systematic way. \mygi\ allows to estimate the effect through a Montecarlo test set. We have again constructed a 1000-events test set but, differently from what done in \ref{montecarlo:sn}, we have now analyzed one single realization of S/N=80 noise injection, starting from the same \object{HD 126681} high-S/N spectrum. The test set has been constructed now by drawing 1000 \Teff\ values from a gaussian distribution centered at 5552 K and with a $\sigma$ of 150 K, and having \mygi\ run with fixed \Teff. The resulting histograms are shown in Fig. \ref{histo:phototemp}, the resulting distributions in the other parameters and metallicity are again given in Table \ref{montecarlo:table}. Although we do not present it here, a more detailed species-by-species error estimate is of course possible, thus allowing to properly assess parameter-related uncertainties on any abundance. The same kind of procedure is of course possible when more than one parameter is kept fixed (e.g. the case of photometric \Teff\ and \glog\ derived from isochrones).


\section{Performance, system requirements, and availability}
\label{performa}

\mygi\ is written in Fortran 90 with Intel extensions.
So far it has been compiled and run under Linux and Mac OS X. 

\begin{table*}
\caption{Average parameter values and $1 \sigma$ intervals for the Montecarlo tests described in \ref{montecarlo}. {In order, the result of the analysis of the noiseless synthetic spectrum, of the three 1000 noise realization runs on the synthetic (for both see \ref{montecarlo:internal}), of the three 1000 noise realization runs on \object{HD126681}, and of the ``photometric \Teff'' run on \object{HD126681} (for both see \ref{montecarlo:sn}).
In the last row, \Teff\ and its $\sigma$ are in {\bf boldface} to indicate they are not results, but the fixed input \Teff\ distribution}}             
\label{montecarlo:table}      
\centering                          
{
\begin{tabular}{l r l r l r l r l }        
\hline \hline                 
Test type & \Teff\ & $\sigma$ & \glog\ & $\sigma$ & \feh & $\sigma$ & \Vturb\ & $\sigma$ \\
          & K      &          & $cm ~s^{-1}$&        &      &          & $km~s^{-1}$ &    \\   
\hline
synthetic noiseless & 5403 & --  & 4.51 & -- & --1.00 & -- & 0.97 & -- \\ 
\hline
synthetic S/N=80 & 5409 & 31  & 4.51 & 0.06 & --1.01 & 0.02 & 1.05 & 0.07 \\
synthetic S/N=50 & 5419 & 36  & 4.54 & 0.08 & --1.00 & 0.03 & 1.04 & 0.09 \\
synthetic S/N=20 & 5461 & 80  & 4.57 & 0.18 & --0.95 & 0.06 & 0.94 & 0.22 \\
\hline                        
\object{HD126681} S/N=80 & 5536 & 36  & 4.42 & 0.07 & --1.24 & 0.03 & 0.58 & 0.16 \\
\object{HD126681} S/N=50 & 5499 & 46  & 4.34 & 0.10 & --1.24 & 0.03 & 0.47 & 0.19 \\
\object{HD126681} S/N=20 & 5422 & 87  & 4.10 & 0.20 & --1.22 & 0.07 & 0.39 & 0.28 \\
\hline
\object{HD126681} fixed \Teff, S/N=80       & {\bf 5552} & {\bf 150} & 4.43 & 0.29 & --1.24 & 0.12 & 0.77 & 0.22 \\
\hline 
\end{tabular}
}
\end{table*}

\mygi\ synthetic grids can grow to significant sizes when covering large spectral ranges and a large parameter space. As an example we can consider the grid employed in Sections \ref{stars:12}, \ref{stars:14}, and \ref{montecarlo}, intended to handle UVES RED 580 spectra of metal poor, cool turn-off and dwarf stars. It is composed of two frames (460 to 590, and 560 to 690 nm) synthesized \citep[from ATLAS12 atmosphere models and by using SYNTHE, see][]{sbordone04,castelli05,kurucz05,sbordone05} with velocity-equispaced, R=300000 sampling. The spectra are computed for 3840 parameter values\footnote{Listing start, end, step and number of steps. \Teff[K]: 5000, 6000, 200, 6; \glog[$cm s^{-1}$]: 3.0, 5.0, 0.5, 5; \Vturb[$km s^{-1}$]: 0.0, 3.0, 1.0, 4; \feh: --4.0, --0.5, 0.5, 8; \alphafe:--0.4, 0.8, 0.4, 4.}. The binary packaged grid has a size of about 2.2GB.
The size of the grid sets the main resource requirement, since the grid is twice resident in memory (the master copy read in at the beginning of the run, and the one, smaller, resampled to the observed pixels of the star being currently processed). Also, grid resampling at observed spectrum pixel wavelengths is usually the longest operation in the processing of a star. For both these reasons, it is always advisable to use the smallest desirable grid both in terms of parameter space and spectrum coverage. 

As a benchmarking and resource consumption reference we use typical values for one of the S/N Montecarlo runs described in \ref{montecarlo:sn}. The run was executed on a quad-core Intel Xeon E31245, 3.3GHz, machine with 11.6 GB of RAM, running openSUSE 12.3 with 3.7.10 64bit kernel. Employing the grid described above in this same section, peak \mygi\ memory consumption was 3.5 GB, with an average processing time per star (over 1000 stars) of 119s, running on a single core\footnote{\mygi is not currently written or compiled to use more than one thread. This would surely speed up the execution significantly, but we do not see it as a priority, since the code is already fast enough that the actual bottleneck is the ``human time'' to prepare the input and check the results, rather than the processor time.}.  These values are to be considered upper limits. In the described case the resampling of $\sim 74700$ synthetic grid ``pixels'' over the $\sim$71000 observed ones takes somewhat more of 50\% of the total per-star processing time, almost all the rest being taken by the $\sim$280 parameter-finding iterations needed to determine \Teff, \glog\, \Vturb, \feh, and \alphafe. At the other extreme, determining only \feh, \alphafe, and detailed abundances at pre-defined atmospheric parameters on data constituted by two FLAMES high-resolution settings takes less than 10 seconds per star, thanks equally to the lack of parameter iteration and to the small number of pixels over which grid resampling should be performed.

For the time being, \mygi\ is available on a collaboration basis only. A website has been set up\footnote{\url{http://mygisfos.obspm.fr}} Eto make detailed code documentation available to interested parties, so that they can assess whether it can be of use to them before contacting us. 

\section{Conclusions}
 
We have presented \mygi, an automatic code for the determination of atmospheric parameters and detailed chemical abundances in cool stars. Replicating as closely as possible a ``traditional'' manual abundance analysis technique for this type of stars, \mygi\ derives results directly and quickly comparable with well known analysis techniques. {Currently, \mygi\ has been employed on the same type of data that are best suited for the manual techniques it replicates, i.e. those characterized by high-resolution and/or broad spectral coverage, such as the ones delivered by UVES, X-Shooter, HARPS, Sophie \citep[][]{caffau11,bonifacio12,caffau12,caffau13,duffau14}}. Its very fast operation (2 minutes or less per star on a mainstream personal computer) is convenient in a number of circumstances. 
\begin{itemize}
\item When analyzing large datasets of (broadly) homogeneous data in the context of large scale spectroscopic surveys such as for instance the Gaia-ESO Public Survey  \citep[{UVES data}, see][]{gilmore12}. 
\item When screening large amounts of low-intermediate resolution spectra for interesting candidates to be followed up. While \mygi\ delivers its full capability for high-resolution (R$\geq$ 25000), large spectral coverage data, it can be effectively used to derive global metallicity from low resolution spectra, once parameters can be inferred from other sources. In this context Abbo \citep[][]{bonifacio03}, the code \mygi\ has been derived from, has been employed to select extremely metal poor Turn-Off candidates from low-resolution SDSS-SEGUE \citep[][]{york00,yanny09} spectra, which were then followed up at high resolution, showing a remarkable selection success rate \citep[e. g. ][]{caffau12,bonifacio12,caffau11}, and \mygi\ is currently used in the same capacity in the context of the TOPoS ESO Large Program \citep[Turn-Off Primordial Stars, ][]{caffau13}.
\item When extending pre-existing literature datasets with new data: when adding new stars to pre-existing samples of abundances derived in previous papers, it is usually impractical to re-analize the existing corpus of data, so homogeneity issues arise because of differences e.g. in atmosphere modeling, or atomic data choices. The very fast processing of \mygi\ allows, on the other hand, to repeat the analysis of the whole sample every time in a fully homogeneous way. As a matter of fact, fully parallel analyses can be conducted e. g. with different atmospheric parameter choices, with practically no time penalty.
\item To consistently tackle error propagation through the analysis pipeline by means of Montecarlo simulations, as presented in Sect. \ref{montecarlo:phototemp} for photometric effective temperatures. This is almost never done due to the very high time cost it would usually imply, but it would be of great importance since errors in atmospheric parameters and abundances are in general strongly correlated, and method dependent.
\end{itemize}

\begin{acknowledgements}
      LS, EC, and SD aknowledge the support of 
      Sonderforschungsbereich SFB 881 ``The Milky Way System'' (subprojects A4 and A5) 
      of the German Research Foundation (DFG).
      PB acknowledges support from the Conseil Scientifique de 
      l'Observatoire de Paris and from the Programme National
      de Cosmologie et Galaxies of the Institut National des Sciences
      de l'Univers of CNRS.
      This research has made use of the SIMBAD database,
	  operated at CDS, Strasbourg, France, and of NASA's Astrophysics Data System, and is partly 
	  based on data obtained from the ESO Science Archive Facility.
\end{acknowledgements}


\end{document}